\documentclass[prl,aps,twocolumn]{revtex4}
\usepackage{amsmath}
\usepackage{amssymb}
\usepackage{graphicx}
\usepackage{subfigure}
\usepackage{algorithm}
\usepackage{bm}
\usepackage{color}

\usepackage{commath}
\usepackage{dcolumn}
\usepackage{multirow}

\usepackage{hyperref}

\begin{document}

\title{
 Kinked Entropy and Discontinuous Microcanonical Spontaneous Symmetry Breaking
}

\author{Hai-Jun Zhou$^{1,2}$}

\affiliation{
  $^1$CAS Key Laboratory for Theoretical Physics, Institute of Theoretical Physics, Chinese Academy of Sciences, Beijing 100190, China\\
  $^2$School of Physical Sciences, University of Chinese Academy of Sciences, Beijing 100049, China
}
\date{\today}

\begin{abstract}
  Spontaneous symmetry breaking (SSB) in statistical physics is a macroscopic collective phenomenon. For the paradigmatic $Q$-state Potts model it means a transition from the disordered color-symmetric phase to an ordered phase in which one color dominates. Existing mean field theories imply that SSB in the microcanonical statistical ensemble (with energy being the control parameter) should be a continuous process. Here we study microcanonical SSB on the random-graph Potts model, and discover that the entropy is a kinked function of energy. This kink leads to a discontinuous phase transition at certain energy density value, characterized by a jump in the density of the dominant color and a jump in the microcanonical temperature. This discontinuous SSB in random graphs is confirmed by microcanonical Monte Carlo simulations, and it is also observed in bond-diluted finite-size lattice systems.
\end{abstract}

\maketitle

Spontaneous symmetry breaking (SSB) is a fundamental concept of physics and is tightly linked to the origin of mass in particle physics, the emergence of superconductivity in condensed-matter system, and the ferromagnetic phase transition in statistical mechanics, to name just a few eminent examples~\cite{Brading-etal-2017}. In statistical physics a theoretical paradigm for SSB is the Potts model, a simple two-body interaction graphical system in which each vertex has $Q$ discrete color states~\cite{Potts-1952,Wu-1982,Baxter-1982}. The equilibrium SSB transition of the Potts model in the canonical ensemble, where inverse temperature $\beta$ is the control parameter, has been extensively investigated (see Refs.~\cite{Gorbenko-etal-2018,Blote-etal-2017,Hu-Deng-2015,Wang-Xie-etal-2014,Lee-Lucas-2014,Tian-etal-2013,Chen-etal-2011,Deng-etal-2011} for some of the recent results). Driven by energy--entropy competitions, this transition is a discontinuous phenomenon when $Q$ is sufficiently large, with the density $\rho_1$ of the dominant color jumps from $1/Q$ to a much larger value at the critical inverse temperature $\beta_{{\rm c}}$. To compensate for the extensive loss of entropy, such a discontinuous transition is always accompanied by a discontinuous decrease of the system's energy density $u$~\cite{Wu-1982,Baxter-1982}.

When the system is isolated and cannot exchange energy with the environment (the microcanonincal ensemble~\cite{Gross-etal-1996,Gross-1997,MartinMayor-2007,Moreno-etal-2018}), it is generally believed that the SSB transition will occur gradually, with the dominant color density $\rho_1$ deviating from $1/Q$ continuously at certain critical energy density $u_{{\rm mic}}$. Indeed if the microscopic entropy density $s(u)$ is a $C^1$-continuous function of energy density $u$ (i.e., both $s(u)$ and its first derivative are continuous), there is no reason to expect a discontinuity of the order parameter $\rho_1$. The $C^1$-continuity of $s(u)$ can be easily verified for the mean field Potts model on a complete graph~\cite{SInote2019}. For finite-dimensional lattices the phase separation mechanism (the nucleation and expansion of droplets~\cite{Biskup-etal-2002,Binder-2003,MacDowell-etal-2006,Nogawa-etal-2011}) will guarantee a $C^1$-continuous entropy profile in the thermodynamic limit. For random graph systems one would na\"ively expect $u_{{\rm mic}}$ to be an inflection point of $s(u)$~\cite{Xu-etal-2018}, which ensures $C^1$-continuity.

\begin{figure}[b]
  \centering
  \includegraphics[width=0.5\linewidth]{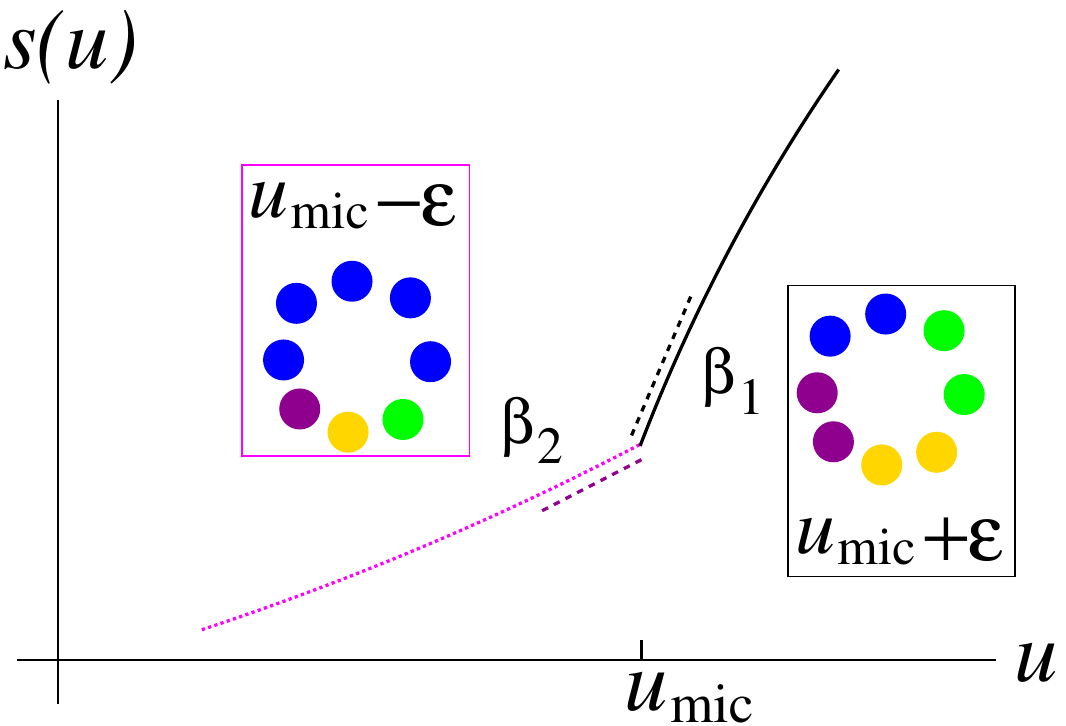}
  \caption{\label{fig:PottsKink}
    Schematic drawing of kinked entropy density $s(u)$. As the energy density $u$ of the $Q$-state Potts model decreases to $u_{{\rm mic}}$, $s(u)$ changes from concave to convex and its slope drops from $\beta_1$ to $\beta_2$. The system is color-symmetric at $u_{{\rm mic}}\! +\! \varepsilon$ ($\varepsilon\! \rightarrow\! 0$) with a lower microcanonical temperature $1/ \beta_1$, but at $u_{{\rm mic}}\! -\! \varepsilon$ it has a highly dominant color and a higher microcanonical temperature $1/\beta_2$.
  }
 \end{figure}
 
In this Letter we investigate the microcanonical Potts model on random graphs using the Bethe-Peierls mean field theory, and discover that the entropy density $s(u)$ is actually not $C^1$-continuous but is kinked at $u\! =\! u_{{\rm mic}}$ for any $Q\! \geq\! 3$ (Fig.~\ref{fig:PottsKink}). Consequently, there is a discontinuous microcanonical phase transition at $u_{{\rm mic}}$, with a jump in the dominant density $\rho_1$ and a drop in the microcanonical inverse temperature. This SSB transition is driven completely by entropy competitions between the microcanonical polarized (MP) phase and the disordered symmetric (DS) phase, and at $u_{{\rm mic}}$ the MP phase is hotter than the DS phase. These theoretical predictions for random graphs are verified by microcanonical Monte Carlo simulations. The discontinuous SSB transition is also observed in three- and higher-dimensional bond-diluted lattices, but only for system sizes not too large~\cite{SInote2019}. The phenomenon of kinked entropy may persist in other multiple-state spin glass systems or combinatorial optimization problems~\cite{Mezard-Montanari-2009}. Our work also adds new insight on the debate about ensemble inequivalence~\cite{Mukamel-2008,Campa-etal-2009,Murata-Nishimori-2012,Touchette-2015}.

\emph{Mean field theory.}--- Consider a graph $G$ formed by $N$ vertices and $M$ edges. Each vertex $i$ has a discrete color $c_i \in \{1, 2, \ldots, Q\}$ and an edge $(i, j)$ between vertices $i$ and $j$ has a ferromagnetic interaction energy $E_{i j}(c_i, c_j) = - \delta_{c_i}^{c_j}$, where $\delta_{c_i}^{c_j}\!=\! 1 (0)$ if $c_i\! =\! c_j$ ($c_i\! \neq\! c_j$).  The total energy of a color configuration $\bm{c} \equiv (c_1, c_2, \ldots, c_N)$ is the summed edge energies, $E(\bm{c}) = \sum_{(i, j)\in G}  E_{i j}(c_i, c_j)$, which is symmetric with respect to color permutations. The partition function $Z(\beta)$ at a given inverse temperature $\beta$ is
\begin{equation}
  \label{eq:Zpotts}
  Z(\beta) \equiv \sum\limits_{\bm{c}} e^{-\beta E(\bm{c})}
  = \sum\limits_{\bm{c}}
  \prod\limits_{(i, j)\in G} \Bigl[1 + (e^\beta - 1) \delta_{c_i}^{c_j}\Bigr] \; .
\end{equation}

We now review the Bethe-Peierls theory for this model~\cite{Huang-1987,Mezard-Montanari-2009}. For simplicity we describe the theoretical equations for random regular (RR) graphs, which are maximally random except that every vertex has exactly $K$ attached edges. (The mean field theory for general graphs can easily be derived following the cavity method of statistical physics~\cite{Mezard-Montanari-2009,Mezard-etal-1987} or through loop expansion of the partition function~\cite{Xiao-Zhou-2011,Zhou-Wang-2012}.) This theory is exact for tree graphs, and because random graphs are locally tree-like (loop lengths diverge logarithmically with $N$) and there is no intrinsic frustration in the edge interactions, we expect it to be exact for RR graphs as well.

Without loss of generality we assume $c\! =\! 1$ to be the dominant (most abundant) color. To compute the marginal probability $\rho_1$ of this color state for a randomly chosen vertex $i$ we first delete $i$ and all its attached edges from the graph. Because short loops are rather rare in the graph, the $K$ nearest neighbors of $i$ will now be far separated in the perturbed cavity graph and consequently their color states will be independent. We denote by $q$ ($\geq 1/Q$) the probability of such a neighboring vertex $j$ to be in state $c_j\! =\! 1$ in the perturbed graph, and assume that vertex $j$ has equal probability $(1-q)/(Q-1)$ to be in any of the other color states.  When vertex $i$ and its $K$ edges are added back to the graph, its probability of being in state $c_i=1$ is then
\begin{equation}
  \rho_1 =
  \biggl[1+ (Q-1) \Bigl(\frac{1+(e^\beta -1) \frac{1-q}{Q-1}}{1+
        (e^\beta - 1) q}\Bigr)^{K} \biggr]^{-1} \; .
\end{equation}
This quantity $\rho_1$ is also the dominant color density of the RR graph. A similar expression for the cavity probability $q$ of the neighboring vertex $j$ can be written down ($j$ has $K\! -\! 1$ edges in the cavity graph):
\begin{equation}
  \label{eq:RRbp}
  q = B(q) \equiv
  \biggl[1+ (Q-1) \Bigl(\frac{1+(e^\beta -1) \frac{1-q}{Q-1}}{1+
   (e^\beta - 1) q}\Bigr)^{K-1} \biggr]^{-1} \; .
\end{equation}
This self-consistent expression is referred to as a belief-propagation (BP) equation~\cite{Pearl-1988}.

The free energy density $f \equiv -(1/ N\beta) \ln Z(\beta)$ of the system can be computed by first summarizing the individual contributions of all the vertices, and then subtracting the individual contributions of all the edges (because each edge contributes to the free energies of two vertices)~\cite{Mezard-Montanari-2009,Mezard-etal-1987,Xiao-Zhou-2011,Zhou-Wang-2012}. At a BP fixed point the explicit expression of $f$ is
\begin{eqnarray}
  f  & = &   -\frac{1}{\beta} \ln \Bigl\{    \bigl[ 1 + (e^\beta -1) q\bigr]^{K}
    \nonumber \\
    & & \quad \quad \quad \quad + (Q-1) \bigl[ 1+ (e^\beta -1)
    \frac{1-q}{Q-1}\bigr]^{K} \Bigr\}
  \nonumber \\
  & &
  + \frac{K}{2 \beta} \ln \Bigl[ 1 + (e^\beta -1) \bigl( q^2 +
    \frac{(1-q)^2}{Q-1}\bigr) \Bigr] \; .
  \label{eq:RRf}
\end{eqnarray}
One can verify that $\frac{\partial f}{\partial q} =0$ when $q=B(q)$. The mean energy density $u$ is obtained from Eq.~(\ref{eq:RRf}) as
\begin{equation}
  u \equiv \frac{\partial (\beta f)}{\partial \beta}
  = - \frac{K}{2} \frac{e^\beta \bigl(q^2 + \frac{(1-q)^2}{Q-1}\bigr)}
  {1 + (e^\beta - 1) \bigl(q^2 + \frac{(1-q)^2}{Q-1}\bigr) } \; .
\end{equation}
The entropy density $s$ of the system is then determined by $s = \beta ( u - f)$~\cite{Huang-1987}.

The BP equation (\ref{eq:RRbp}) always has a trivial fixed point $q\! =\! 1/Q$ which corresponds to the disordered symmetric (DS) phase with all the colors being equally abundant~\cite{SInote2019}. This fixed point becomes unstable with respect to the iteration $q^{t+1}\! \leftarrow\! B(q^t)$ when $\beta \! >\! \beta_{{\rm DS}}\! \equiv\! \ln\bigl(1+Q/(K-2)\bigr)$. For $K\! \geq\! 3$ and $Q\! \geq\! 3$, Eq.~(\ref{eq:RRbp}) has a stable fixed point with $q$ and $\rho_1$ strictly larger than $1/Q$ at $\beta \! >\! \beta_{{\rm CP}}$, which corresponds to the canonical polarized (CP) phase of broken color symmetry. Here $\beta_{{\rm CP}}$ ($< \beta_{{\rm DS}}$) is the lowest inverse temperature at which the CP phase becomes possible. The CP and DS phases have equal free energy density at a critical inverse temperature $\beta_{{\rm c}} \in (\beta_{{\rm CP}}, \beta_{{\rm DS}})$, so an equilibrium phase transition occurs at $\beta_{{\rm c}}$, with a sudden drop in energy density $u$~\cite{SInote2019}.

\begin{figure*}[t]
  \centering
  \subfigure[]{
    \label{fig:RRK4Q6Beta:b}
    \includegraphics[angle=270, width=0.35\linewidth]{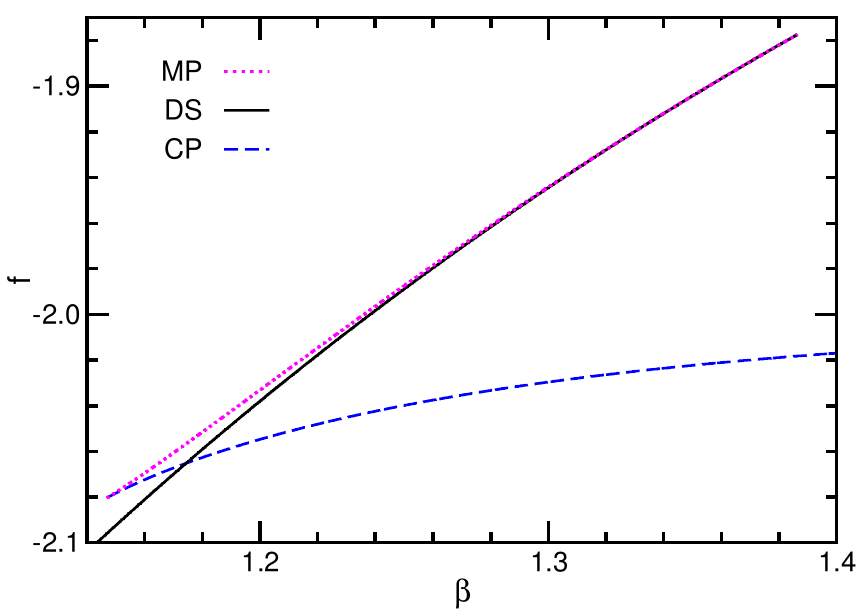}
  }
  \subfigure[]{
    \label{fig:RRK4Q6Beta:c}
    \includegraphics[angle=270, width=0.35\linewidth]{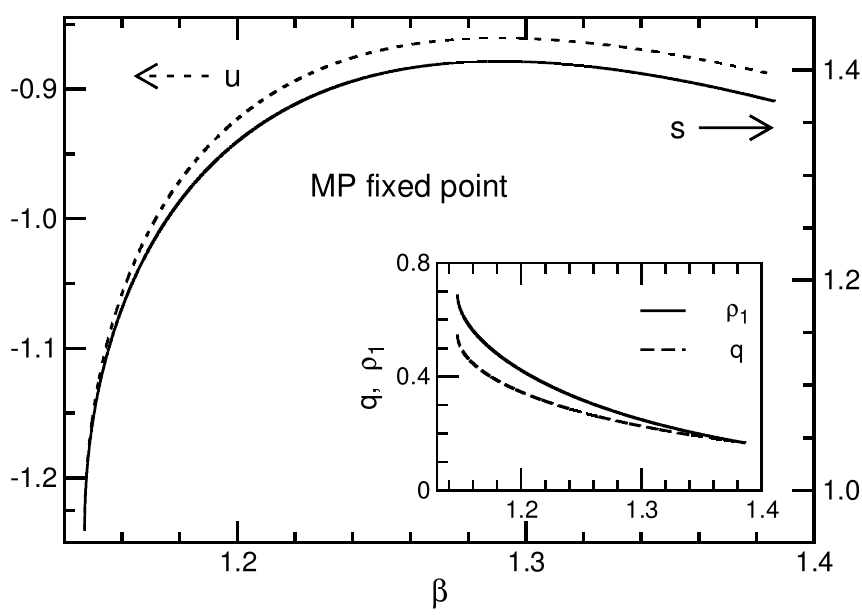}
  }
  
  \subfigure[]{
    \label{fig:RRK4Q6Micro:a}
    \includegraphics[angle=270, width=0.35\linewidth]{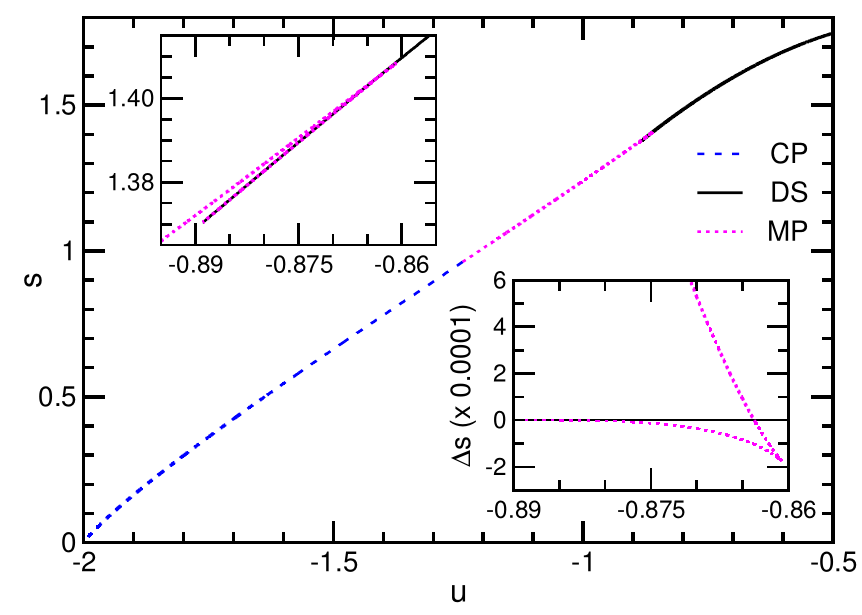}
  }
  \subfigure[]{
    \label{fig:RRK4Q6Micro:b}
    \includegraphics[angle=270, width=0.35\linewidth]{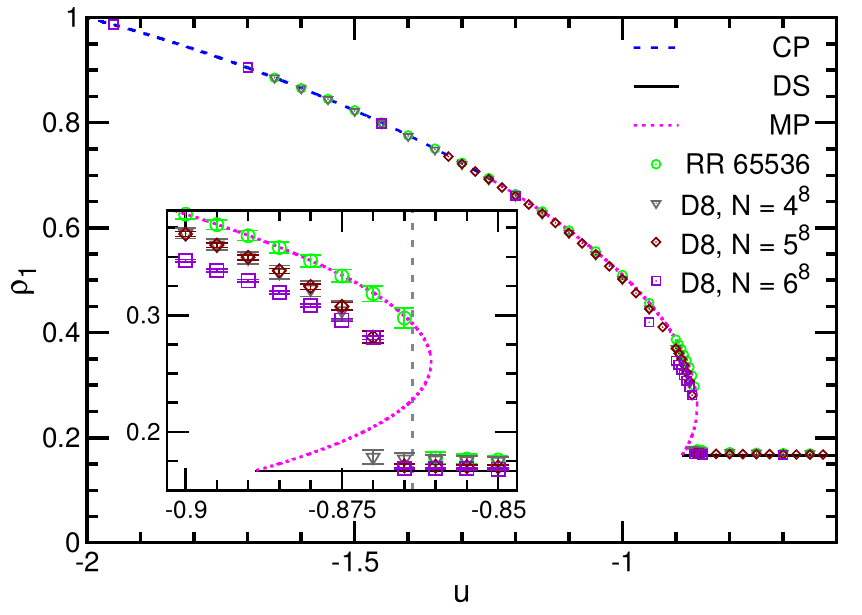}
  }
  \caption{
    \label{fig:RRK4Q6Beta}
    Potts model on regular random graphs, $K\! =\! 4$ and $Q\! =\! 6$. (a) Free-energy densities $f(\beta)$ for the disordered symmetric (DS, solid line), the canonical polarized (CP, dashed line), and the microcanonical polarized (MP, dotted line) fixed points of the BP equation.  The DS solution is stable at inverse temperature $\beta\! <\! \beta_{{\rm DS}}\! =\! 1.386$, the CP solution exists for $\beta\! \geq\! \beta_{{\rm CP}} \! =\! 1.147$, and the DS--CP phase transition occurs at $\beta_{{\rm c}}\! =\! 1.174$ with the energy density $u$ dropping from $-0.786$ to $-1.523$. (b) Energy density $u(\beta)$, entropy density $s(\beta)$, fixed-point value $q(\beta)$ and density $\rho_1(\beta)$ of the dominant color (inset), for the MP fixed point. The maximal achievable MP energy density is $u_{{\rm max}}\! = \! -0.861$. (c) and (d): Entropy density $s$ and dominant color density $\rho_1$ versus energy density $u$ for the DS (solid line), MP (dashed line), and CP (dotted line) fixed points. Upper-left and lower-right insets of (c) show an enlarged view of the MP entropy profile and the difference $\Delta s$ between the MP and DS entropy density values ($\Delta s \! = \! 0$ at energy density $u_{{\rm mic}}\! =\! -0.864$). Symbols in (d) are microcanonical Monte Carlo simulation results obtained on a single RR graph ($N\! =\! 65536$) and several bond-diluted eight-dimensional periodic hypercubic lattices of side length $L=4, 5, 6$ (D8, $N\! =\! L^8$), degree $K\! =\! 4$ and $Q\! =\! 6$. The inset of (d) is an enlarged view of the transition region, and the phase transition point $u_{{\rm mic}}$ for RR graphs is marked by the vertical dashed line, at which $\rho_1$ jumps from $1/6$ to $0.293$.
  }
\end{figure*}

\emph{Microcanonical SSB.}--- For $\beta \in (\beta_{{\rm CP}}, \beta_{{\rm DS}})$ the BP equation (\ref{eq:RRbp}) has another fixed point which is unstable with respect to $q^{t+1}\leftarrow B(q^t)$~\cite{SInote2019}. This fixed point is usually neglected because its free energy is higher than those of the DS and CP phases (Fig.~\ref{fig:RRK4Q6Beta:b}). But we find that it reveals a discontinuous microcanonical phase transition between the DS phase and a new microcanonical polarized (MP) phase of the configuration space.

Plotting the predicted thermodynamic values of the MP fixed point (Fig.~\ref{fig:RRK4Q6Beta:c}), we observe that while $q$ and $\rho_1$ are monotonic functions of $\beta$ as anticipated, the energy density $u$ and entropy density $s$ both are non-monotonic. This surprising feature of $u$ and $s$ leads to the two-branched entropy profile shown in the upper-left inset of Fig.~\ref{fig:RRK4Q6Micro:a}. These two entropy branches merge and stop at $u_{{\rm max}}$, which is the maximal achievable energy density of the MP phase. The entropy of the lower MP branch is slightly lower than that of the DS phase so this branch has no physical significance. On the other hand, the entropy of the upper MP branch exceeds that of the DS phase as $u$ decreases below certain critical value $u_{{\rm mic}}$ which is strictly lower than $u_{{\rm max}}$, indicating the system will jump from the color-symmetric phase to a color-symmetry-broken MP phase which is stable only in the microcanonical ensemble. The dominant color density $\rho_1$ at $u_{{\rm mic}}$ is strictly higher than $1/Q$, so the spontaneous breaking of color symmetry is a discontinuous emerging phenomenon. Notice that at $u$ slightly below $u_{{\rm mic}}$ the entropy density of the MP phase is higher than that of the DS phase.

Because the entropy densities of the DS and MP phases are equal at $u\! =\! u_{{\rm mic}}$ but have different slopes (Fig.~\ref{fig:RRK4Q6Micro:a}), the system's entropy density function $s(u)$ is not $C^1$-continuous but is kinked at $u_{{\rm mic}}$~\cite{SInote2019}.  Since the microcanonical inverse temperature is equal to the first derivative of $s(u)$, $\beta \equiv \frac{{\rm d} s(u)}{{\rm d} u}$~\cite{Huang-1987}, there will be a sudden drop of the microcanonical $\beta$ and an associated sudden drop of the free energy density $f$ ($=u-\beta s$) as the system changes from the DS to the MP phase at $u_{{\rm mic}}$. In other words, at $u_{{\rm mic}}$ the partially ordered MP phase is hotter than the disordered symmetric phase and has a lower free energy density. This peculiar feature of $s(u)$ is qualitatively different from the recently discussed entropy inflection phenomenon, which is associated with the vanishing of the second-order derivative of $s(u)$~\cite{Xu-etal-2018}.

We have checked that as long as $Q\! \geq\! 3$, the discontinuous SSB phenomenon holds for all the RR graph ensembles of degree $K\geq 3$. As demonstrated in Table~\ref{tab:RRresults}, at each fixed value of $Q$ the $\rho_1$ and $\beta$ gaps at $u_{{\rm mic}}$ both decrease with degree $K$ (and vanish gradually as $K\! \rightarrow\! \infty$~\cite{SInote2019}). The discontinuous microcanonical phase transition will also occur in an extended Potts model with additional kinetic energies~\cite{SInote2019}.

\begin{table}
  \caption{\label{tab:RRresults}
    The critical energy density $u_{{\rm mic}}$, the jump $\Delta \rho_1$ of the dominant color density and the drop $\Delta \beta$ of the microcanonical inverse temperature at $u_{{\rm mic}}$, for the $Q$-state Potts model on RR graphs of degree $K$.
  }
  \centering
  \begin{tabular}{c|c|ccc||c|c|ccc}
    $K$ & $Q$ & $u_{{\rm mic}}$ & $\Delta \rho_1$ & $\Delta \beta$ &  $K$ & $Q$ & $u_{{\rm mic}}$ & $\Delta \rho_1$ & $\Delta \beta$ \\ \hline
     \multirow{8}{*}{3} & $3$ & $-0.997$ & $0.065$ & $-0.012$ & \multirow{8}{*}{5} & $3$ & $-1.248$ & $0.040$ & $-0.007$ \\
     & $4$ & $-0.929$ & $0.109$ & $-0.035$ &  & $4$ & $-1.085$ & $0.071$ & $-0.022$\\
    & $5$ & $-0.884$ & $0.138$ & $-0.058$  &  & $5$ & $-0.984$ & $0.092$ & $-0.040$ \\
     & $6$ & $-0.852$ & $0.158$ & $-0.079$ &  & $6$ & $-0.913$ & $0.108$ & $-0.059$ \\
     & $7$ & $-0.827$ & $0.171$ & $-0.099$ &  & $7$ & $-0.860$ & $0.119$ & $-0.076$ \\
     & $8$ & $-0.807$ & $0.182$ & $-0.116$ &  & $8$ & $-0.819$ & $0.127$ & $-0.093$ \\
     & $9$ & $-0.790$ & $0.189$ & $-0.132$ &  & $9$ & $-0.785$ & $0.133$ & $-0.109$ \\
     & $10$ & $-0.776$ & $0.195$ & $-0.146$ &  & $10$ & $-0.757$ & $0.138$ & $-0.124$ \\
 \hline 
    \multirow{8}{*}{4} & $3$ & $-1.108$ & $0.049$ & $-0.009$ & \multirow{8}{*}{6} & $3$ & $-1.398$ & $0.033$ & $-0.005$ \\
    & $4$ & $-0.991$ & $0.085$ & $-0.028$ &  & $4$ & $-1.192$ & $0.061$ & $-0.018$\\
    & $5$ & $-0.916$ & $0.110$ & $-0.048$  &  & $5$ & $-1.065$ & $0.081$ & $-0.034$ \\
    & $6$ & $-0.864$ & $0.126$ & $-0.069$ & & $6$ & $-0.977$ & $0.095$ & $-0.050$ \\
    & $7$ & $-0.824$ & $0.139$ & $-0.088$ & & $7$ & $-0.911$ & $0.105$ & $-0.067$ \\
    & $8$ & $-0.792$ & $0.147$ & $-0.105$ & & $8$ & $-0.861$ & $0.113$ & $-0.083$ \\
    & $9$ & $-0.766$ & $0.154$ & $-0.121$ & & $9$ & $-0.820$ & $0.119$ & $-0.098$ \\
    & $10$ & $-0.745$ & $0.159$ & $-0.136$ & & $10$ & $-0.786$ & $0.124$ & $-0.112$ \\
 \hline 
  \end{tabular}
\end{table}

\emph{Monte Carlo simulations.}--- We carry out microcanonical Monte Carlo (MC) simulations to check the theoretical predictions. There are many discussions on  microcanonical MC methods~\cite{Creutz-1983,Lee-1995,Gross-etal-1996,MartinMayor-2007,Schierz-etal-2016,Moreno-etal-2018}, and here we employ the simple demon method~\cite{Creutz-1983} to draw a set of independent configurations which are located slightly below a prescribed objective energy level $E_{{\rm o}}$.  Starting from an initial configuration $\bm{c}$ of energy $E\! \leq\! E_{{\rm o}}$, an elementary MC step unfolds as follows: (1) pick a vertex $i$ uniformly at random and change its color $c_i$ to a uniformly random new value $c_i^\prime$ ($\neq \! c_i$); (2) accept this color change if the energy $E^\prime$ of the resulting new configuration satisfies $E^\prime \leq E_{{\rm o}}$, otherwise keep the old color $c_i$; (3) increase the evolution time $t$ by a tiny amount $1/N$ (one unit time therefore corresponds to $N$ single-flip trials). This MC dynamics obeys detailed balance, so the sampled color configurations all have the same statistical weight. The simulation results obtained on a large RR graph instance are shown in Fig.~\ref{fig:RRK4Q6Micro:b} ($K\! = \! 4, Q\! = \! 6$). We indeed observe a discontinuous transition at the predicted critical energy density $u_{{\rm mic}}$. The numerical results on the dominant color density $\rho_1$ also agree perfectly with theory. The predicted inverse temperature gap $\Delta \beta$ is also quantitatively confirmed by computer simulations~\cite{SInote2019}.

We also consider bond-diluted $D$-dimensional hypercubic lattices of side length $L$ with periodic boundary conditions ($N\! =\! L^D$). By keeping only $K$ bonds in a maximally random manner for every vertex (see \cite{SInote2019} for construction details), the shortest loops passing through the vertices rapidly increase their lengths as $K$ decreases and $D$ increases, and the diluted lattice is locally resembling a random graph~\cite{Fernandez-etal-2010}. A discontinuous SSB transition is observed in the MC dynamics for such bond-diluted lattice systems at high dimensions (e.g., $D\! =\! 8$, Fig.~\ref{fig:RRK4Q6Micro:b}) and also at the physical dimension $D\! = \! 3$~\cite{SInote2019}. However, unlike the case of truly random graphs, we expect that the SSB transition in these lattice systems will become continuous in the thermodynamic limit~\cite{SInote2019}, because phase separation is deemed to occur as the system size $L$ becomes sufficiently large~\cite{Biskup-etal-2002,Binder-2003,MacDowell-etal-2006,Nogawa-etal-2011}.

\emph{Conclusion.}--- In summary, we predicted and confirmed a discontinuous microcanonical SSB phase transition in the $Q$-state Potts model on random graphs. Such a discontinuous transition was also observed in bond-diluted finite-size lattice systems (even down to three dimensions~\cite{SInote2019}).  In the future we need to investigate the geometric property of the configurations in the MP phase (e.g., the possibility of a percolating cluster of connected same-color vertices)~\cite{MacDowell-etal-2006}, and possible latent structures prior to the microcanonical transition~\cite{Zhou-Ma-2009,Zhou-Wang-2010}, and to study systematically the microcanonical SSB transition in finite-dimensional finite-size systems and the associated inequivalence between the microcanonical and the canonical ensembles~\cite{Mukamel-2008,Campa-etal-2009,Murata-Nishimori-2012,Touchette-2015}. The discovered property of kinked entropy may be a general feature of random graphical models with a canonical discontinuous phase transition, and it may have important computational consequences in optimization tasks~\cite{Mezard-Montanari-2009}.

\begin{acknowledgments}
  The following funding supports are acknowledged: National Natural Science Foundation of China Grants No. 11421063 and No. 11747601; the Chinese Academy of Sciences Grant No. QYZDJ-SSW-SYS018. Numerical simulations were carried out at the HPC cluster of ITP-CAS and also at the Tianhe-2 platform of the National Supercomputer Center in Guangzhou. The author thanks Youjin Deng, Gaoke Hu, Hao Hu, Shaomeng Qin, Mutian Shen, and Jinhua Zhao for helpful discussions and/or valuable comments on the manuscript. 
\end{acknowledgments}



\clearpage

\widetext

\begin{appendix}

\begin{center}
  {\bf{Kinked Entropy and Discontinuous Microcanonical Spontaneous Symmetry Breaking}}
  \vskip 0.1cm
  Hai-Jun Zhou
  \vskip 0.3cm
  Supplementary Information
\end{center}

\vskip 1cm

\section*{S1: Exact results on complete graphs}

Consider a complete graph in which every vertex interacts with every other vertex. The energy of a color configuration $\bm{c}=(c_1, c_2, \ldots, c_N)$ is
\begin{equation}
  E(\bm{c}) = - \frac{1}{N} \sum\limits_{i=1}^{N-1} \sum\limits_{j=i+1}^{N}
  \delta_{c_i}^{c_j} \; ,
\end{equation}
where the rescaling factor $\frac{1}{N}$ is introduced to make the total energy an extensive quantity. Suppose there are $N_c$ vertices of color $c$ in the configuration $\bm{c}$, then the total energy can be rewritten as
\begin{equation}
  E(\bm{c})  =  \frac{1}{2} - \frac{N}{2}
  \sum\limits_{c=1}^Q \Bigl(\frac{N_c}{N}\Bigr)^2
  =  \frac{1}{2} -\frac{N}{2} \sum\limits_{c=1}^{Q} \rho_c^2 
  \; ,
\end{equation}
where $\rho_c \equiv \frac{N_c}{N}$ is the density of color $c$. In the thermodynamic limit of $N\! \rightarrow\! \infty$ the energy density $u$ is simply
\begin{equation}
  u = - \frac{1}{2} \sum\limits_{c=1}^{Q} \rho_c^2 \; .
\end{equation}
The total number of microscopic configurations corresponding to the coarse-grained state $(N_1, N_2, \ldots, N_Q)$ is $\Omega(N_1, N_2, \ldots, N_Q) = \frac{N!}{N_1! N_2! \ldots N_q!}$. In the thermodynamic limit then the entropy density $s$ is
\begin{equation}
  s = -\sum\limits_{c=1}^{N} \rho_c \ln \rho_c \; .
\end{equation}

The task is now to find the values of $(\rho_1, \rho_2, \ldots, \rho_Q)$ which lead to the maximum of $s$ under the constraints of fixed energy density $u$ and fixed number $N$ of vertices. This can be achieved by introducing a function $z(\rho_1, \ldots, \rho_Q)$ with two Lagrange multipliers $\lambda_1$ and $\lambda_2$:
\begin{equation}
  z  =
  - \sum\limits_{c=1}^{N} \rho_c \ln \rho_c
  + \frac{\lambda_1}{2} \sum\limits_{c=1}^{Q} \rho_c^2
  + \lambda_2 \sum\limits_{c=1}^{N} \rho_c \; .
\end{equation}
The first derivative of this function with $\rho_c$ is $\frac{\partial z}{\partial \rho_c} = - \ln \rho_c -1 + \lambda_1 \rho_c + \lambda_2$. Therefore, from the condition $\frac{\partial z}{\partial \rho_c} = 0$ we obtain that
\begin{equation}
  \label{eq:xcfull}
  \rho_c =  \frac{e^{\lambda_1 \rho_c}}{\sum_{c^\prime = 1}^Q e^{\lambda_1 \rho_{c^\prime}}}
  \quad\quad\quad (c=1, 2, \ldots, Q) \; .
\end{equation}

\begin{figure}[t]
  \centering
  \includegraphics[angle=270, width=0.475\linewidth]{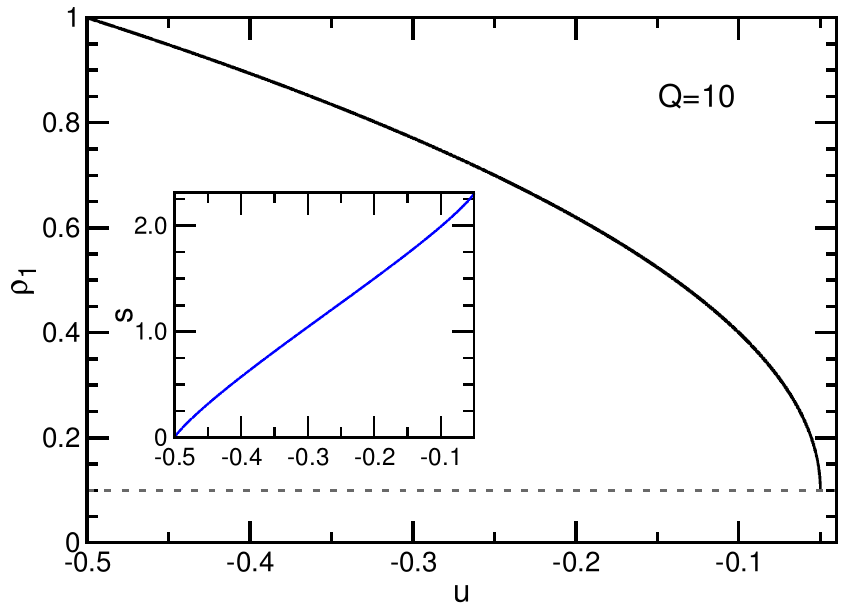}
  \caption{
    \label{fig:CompQ10}
    $Q$-state Potts model on a complete graph. The maximum energy density is $u\! =\! -\frac{1}{2 Q}$ and the minimum energy density is $u\! =\! -\frac{1}{2}$. There is a continuous microcanonical SSB transition at energy density $u_{{\rm mic}}\! =\! -\frac{1}{2 Q}$, which is identical to the maximum energy density ($u_{{\rm mic}}\! =\! - 0.05$ in the example of $Q\! =\! 10$ shown here). The density $\rho_1$ of the dominant color deviates continuously from $\frac{1}{Q}$ (marked by the horizontal dashed line). The inset shows the non-concave relationship between the entropy density $s$ and the energy density $u$.
  }
\end{figure}

The color-symmetric fixed-point solution of Eq.~(\ref{eq:xcfull}) is $\rho_c \! =\! \frac{1}{Q}$ for all colors $c$. The energy density of this disordered symmetric (DS) solution is the maximum value $u\! =\! -\frac{1}{2 Q}$, and its entropy density is $\ln Q$. If the energy density $u$ decreases from the maximum value, then color symmetry has to be broken. Therefore the critical energy density for spontaneous symmetry breaking (SSB) is simply $u_{{\rm mic}}= - \frac{1}{2 Q}$.

The other fixed-point solutions of Eq.~(\ref{eq:xcfull}) can be characterized by two parameters, $\rho_1$ and $m$. The real parameter $\rho_1 \in [\frac{1}{Q}, 1]$ is the density of a dominant color, and the integer $m\in\{1, 2, \ldots, Q-1\}$ is the number of dominant colors. Without loss of generality we assume that $\rho_i=\rho_1$ for $i=1, 2, \ldots, m$ and $\rho_{j}=\frac{1-m \rho_1}{Q-m}$ for $j=m+1, m+2, \ldots, Q$. At a fixed integer value of $m$, the order parameter $\rho_1$ is expressed as
\begin{equation}
  \rho_1 =
  \frac{1}{Q} + \sqrt{2 \bigl(\frac{1}{m}-\frac{1}{Q}\bigr) (u_{{\rm mic}}-u)} \; .
  \label{eq:rho1full}
\end{equation}
Notice that $\rho_1$ is a continuous function of energy density $u$, so the SSB transition at $u_{{\rm mic}}$ must be a continuous phase transition (Fig.~\ref{fig:CompQ10}).

The entropy density $s$ at the polarized fixed point of Eq.~(\ref{eq:xcfull}) is
\begin{equation}
  s = - m \rho_1 \ln \rho_1 - (1- m \rho_1) \ln \frac{1-m \rho_1}{Q-m} \; ,
\end{equation}
where $\rho_1$ is determined by Eq.~(\ref{eq:rho1full}). The parameter $m$ should be set to an integer value which maximizes $s$. It turns out that $m\! =\! 1$ for all values of $Q$. Therefore, in the SSB phase there is only one dominant color, and all the other colors are equally abundant in the system. We find that in the general case of $Q\! \geq\! 3$ the entropy density function $s(u)$ is convex in the vicinity of $u_{{\rm mic}}$ (Fig.~\ref{fig:CompQ10}). This non-concave property means that there is a discontinuous SSB phase transition in the canonical ensemble at certain critical value $\beta_{{\rm c}}$ of the inverse temperature.

To summarize, for the complete-graph $Q$-state Potts model ($Q\! \geq\! 3$), the SSB transition is always a discontinuous phase transition in the canonical ensemble but it is always a continuous phase transition in the microcanonical ensemble.

\section*{S2: The random-graph Potts model in the canonical ensemble}

In the canonical ensemble the inverse temperature $\beta$ of the environment is the control parameter, and the energy density $u$ of the system is not fixed. We now briefly describe some of the results obtained by the Bethe-Peierls mean field theory and by canonical Monte Carlo (MC) simulations. For concreteness we consider regular random (RR) graphs of vertex degree $K\! =\! 4$ and set the number of colors to $Q\! =\! 6$, as in the main text.

\begin{figure}
  \centering
  \subfigure[]{
    \label{fig:RRK4Q6Can:a}
    \includegraphics[angle=270, width=0.475\linewidth]{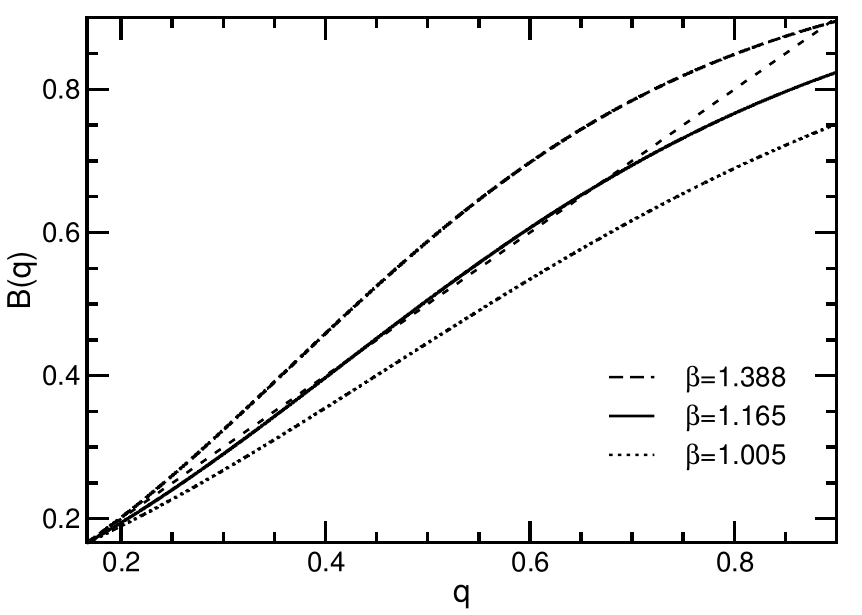}
  }
  
  \subfigure[]{
    \label{fig:RRK4Q6Can:b}
    \includegraphics[angle=270, width=0.475\linewidth]{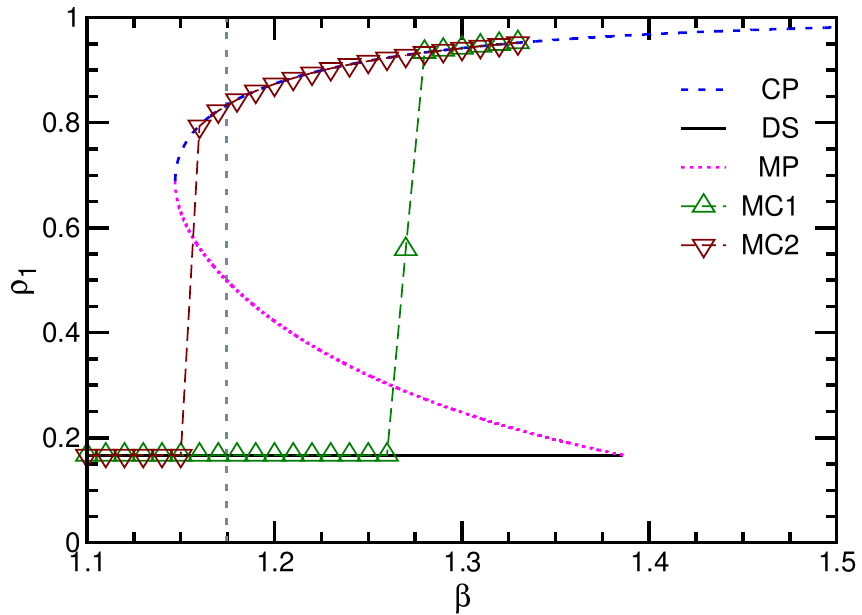}
  }
  \subfigure[]{
    \label{fig:RRK4Q6Can:c}
    \includegraphics[angle=270, width=0.475\linewidth]{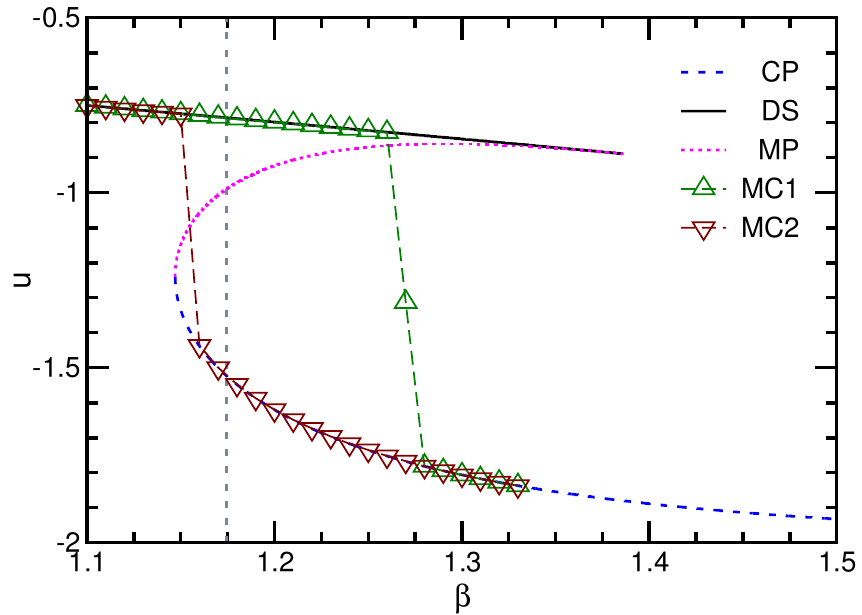}
  }
  \caption{
    \label{fig:RRK4Q6Can}
    $Q$-state Potts model with $Q\! =\! 6$ in the canonical ensemble, on $K\!=\!4$ regular random graphs. (a) The BP fixed points are the intersection points of the curve $B(q)$ and the dashed diagonal line. Depending on $\beta$ there might be one, two, or three fixed points. (b) and (c): The density $\rho_1$ of the dominant color and the mean energy density $u$ for the disordered symmetric (DS, solid line), the canonical polarized (CP, dashed line), and the microcanonical polarized (MP, dotted line) fixed points. The up- and down-triangles are canonical MC simulation results obtained on a single RR graph of size $N=10000$, with the initial color configuration being completely disordered and random (MC1) or being completely ordered (MC2). The vertical dashed lines mark the canonical phase transition point $\beta_{{\rm c}}\! =\! 1.174$.
  }
\end{figure}

First, depending on the value of $\beta$, the BP equation $q = B(q)$ may have one, two, or three fixed-point solutions, see Fig.~\ref{fig:RRK4Q6Can:a}. The trivial fixed point $q\! =\! \frac{1}{Q}$ corresponds to the disordered symmetric (DS) phase and it is locally stable for $\beta\! <\! \beta_{{\rm DS}}\! =\! 1.386$. When $\beta\! \geq\! \beta_{{\rm CP}}\! = \! 1.147$ there is another stable fixed point with $q$ much larger than $\frac{1}{Q}$, which corresponds to the canonical polarized (CP) phase. In this CP phase one color is much more abundant than each of all the other $Q\! -\! 1$ colors (which are equally abundant among themselves), that is, the density $\rho_1$ of the dominant color is much higher than $\frac{1}{Q}$.

The free energy density of the CP phase becomes lower than that of the DS phase as $\beta$ exceeds the critical value $\beta_{{\rm c}}\! = \! 1.174$,  see Fig.~\ref{fig:RRK4Q6Beta:b}. Therefore there is a discontinuous equilibrium phase transition at $\beta_{{\rm c}}$, at which $\rho_1$ jumps from $\frac{1}{Q}\! = \! 0.167$ to a much higher value $0.833$ and $u$ drops from $-0.786$ to $-1.523$. Because of the high free energy barrier between the DS and CP phases, there is a strong hysteresis effect in the canonical MC simulation dynamics in the vicinity of $\beta_{{\rm c}}$,  see Figs.~\ref{fig:RRK4Q6Can:b} and \ref{fig:RRK4Q6Can:c}.

When $\beta\! \in\! (\beta_{{\rm CP}}, \beta_{{\rm DS}})$ the BP equation also has an unstable fixed point, referred to as the microcanonical polarized (MP) one, whose free energy density is higher than those of the DS and CP fixed points. This fixed point therefore is physically irrelevant in the canonical ensemble, see Fig.~\ref{fig:RRK4Q6Can}.

\section*{S3: The entropy kink at $u_{{\rm mic}}$ and the microcanonical inverse temperature}

\begin{figure}
  \centering
  \includegraphics[angle=270,width=0.475\linewidth]{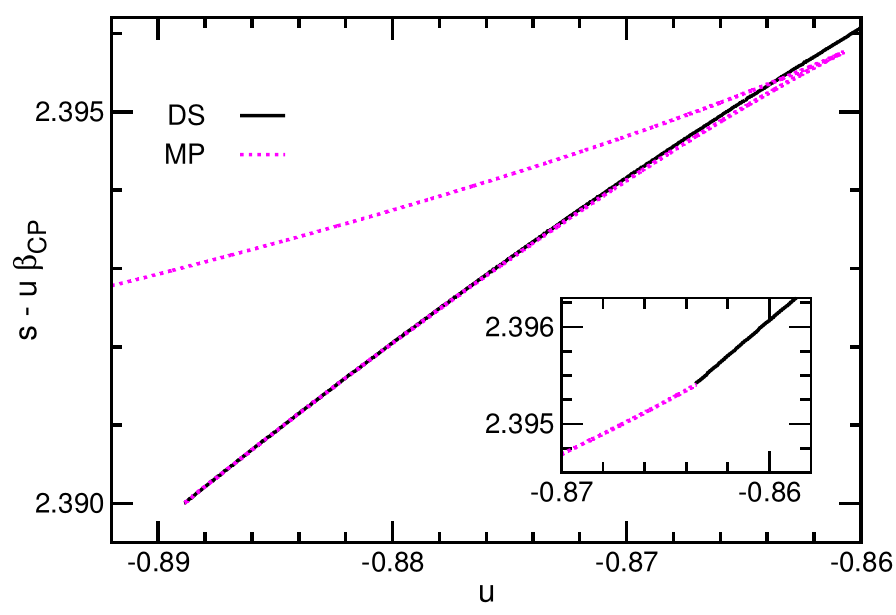}
  \caption{
    \label{fig:k4q6SmEBvsE}
    This figure is complementary to Fig.~2(c) of the main text. The same theoretical data in the upper-left inset of Fig.~2(c) is redrawn here, but with the vertical axis being $\tilde{s}(u) \equiv s(u) - u \beta_{{\rm CP}}$, with $\beta_{{\rm CP}} = 1.147$. The inset here is an enlarged view of the kink of $\tilde{s}(u)$ at $u_{{\rm mic}}\! =\! -0.864$.
  }
\end{figure}

For the Potts model of $Q\! =\! 6$ on the RR graph of degree $K\! =\! 4$, the upper-left inset of Fig.~2(c) in the main text has shown how the entropy densities of the DS and MP fixed-point solutions change with energy density $u$, but the entropy kink is not visually obvious in that figure. To clearly demonstrate the entropy kink, let us define a modified entropy density function $\tilde{s}(u)$ as
\begin{equation}
  \tilde{s}(u) \equiv s(u) - u \beta_{{\rm CP}} \; .
\end{equation}
Since $u \beta_{{\rm CP}}$ is linear in $u$, if $\tilde{s}(u)$ has a kink then the entropy density $s(u)$ will also have a kink. We redraw the theoretical data of Fig.~2(c) at the vicinity of the critical energy density $u_{{\rm mic}} \! = \! -0.864$ in Fig.~\ref{fig:k4q6SmEBvsE}. The kink of the entropy density is now quite evident.

Associated with the entropy kink at $u_{{\rm mic}}$ is the discontinuity of the microcanonical inverse temperature $\beta$. To verify this $\beta$ discontinuity by the microcanonical Monte Carlo (MC) simulation dynamics, we notice that the microcanonical inverse temperature can be estimated by
\begin{equation}
  \beta = \log\Bigl( 1 + \frac{1}{\langle E_{{\rm demon}} \rangle}\Bigr) \; ,
\end{equation}
where $\langle E_{{\rm demon}}\rangle$ is the mean energy of the demon (see Ref.~\cite{Creutz-1983}). The non-negative demon energy $E_{{\rm demon}}$ is simply the difference between the objective energy $E_{{\rm o}}$ and the actual energy $E(\bm{c})$ of the color configuration $\bm{c}$, namely $E_{{\rm demon}} = E_{{\rm o}} - E(\bm{c})$.  Therefore $\langle E_{{\rm demon}} \rangle$ is easy to compute through the microcanonical MC evolution process.

\begin{figure}[b]
  \centering
  \includegraphics[angle=270,width=0.475\linewidth]{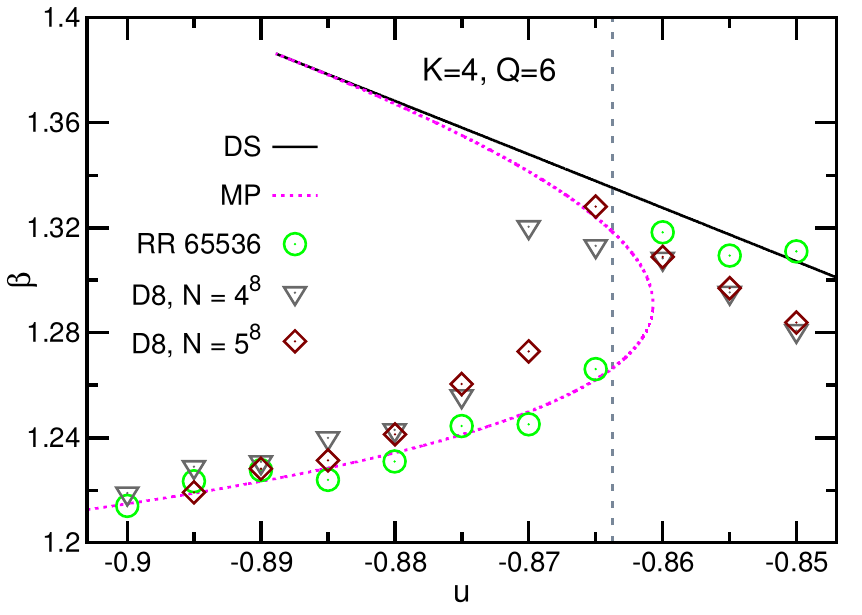}
  \caption{
    \label{fig:k4q6beta}
    The relationship between the microcanonical inverse temperature $\beta$ of the Potts model ($Q\! =\! 6$) and the energy density $u$. Solid line (for the DS phase) and dotted line (for the MP phase) are theoretical predictions for RR graphs of degree $K\! =\! 4$, and the vertical dashed line marks the predicted microcanonical phase transition point $u_{{\rm mic}}$. Symbols are microcanonical Monte Carlo simulation results obtained on the graph instances of Fig.~2(d), including the RR graph ($N\! =\! 65536$) and the two bond-diluted eight-dimensional periodic hypercubic lattices of side length $L=4, 5$ (D8, $N\! =\! L^8$), degree $K\! =\! 4$.
  }
\end{figure}

Figure~\ref{fig:k4q6beta} shows the good agreement between the theoretically predicted and actually measured microcanonical inverse temperatures for the RR graph of degree $K\! = \! 4$ at $Q\! = \! 6$. This figure also shows the measured microcanonical inverse temperatures at different energy densities $u$ for two of the eight-dimensional bond-diluted lattice systems used in Fig.~2(d). An interesting feature is that, given a fixed value of energy density $u$ at the DS phase ($u\! >\! u_{{\rm mic}}$), the microcanonical inverse temperature of $D\!= \!8$ diluted lattice systems is considerably lower than that of the RR graph instance, and the difference increases as $u$ further increases. Further research is needed to fully understand such differences.

\section*{S4: The Potts model on RR graphs of large degree $K$}

We investigate here the asymptotic property of the microcanonical SSB phase transition of the Potts model as the degree $K$ of the RR graphs approaches infinity. For this purpose, it turns out to be convenient to define a parameter, $C_1$, as
\begin{equation}
  C_1 \equiv \frac{e^\beta - 1}{1+ \frac{e^\beta -1}{Q}} \; .
\end{equation}
Let us denote the BP fixed-point solution as
\begin{equation}
  q = \frac{1}{Q} + \Delta \; .
\end{equation}
Then the mean energy density $u$ is expressed as
\begin{equation}
  u = - \frac{\bigl(1+\frac{(Q-1) C_1}{Q}\bigr) K}{2 Q}
  \Bigl(\frac{1+\frac{Q^2}{Q-1} \Delta^2}{1+\frac{Q C_1}{Q-1} \Delta^2}
  \Bigr) \; .
\end{equation}
After some careful derivations, we find that the free energy density difference between the MP phase and the DS phase, at energy density value $u$, is expressed as
\begin{eqnarray}
  s_{{\rm diff}} & = & -\frac{K}{2} \ln \Bigl(1+\frac{Q C_1}{Q-1} \Delta^2 \Bigr)
  + K \ln \bigl( 1 + C_1 \Delta \bigr)
  + \ln\Bigl[ 1 - (1-1/Q)\Bigl(1 - \bigl(1- \frac{(1+1/(Q-1)) C_1 \Delta}
    {1+ C_1 \Delta} \bigr)^K \Bigr)\Bigr] \nonumber \\
  & & + \frac{K}{2} \ln \Bigl(1- \frac{Q^2 \bigl(1+\frac{(Q-1) C_1}{Q}\bigr)}
  {(Q-1)^2 \bigl(1+ \frac{Q C_1}{Q-1} \Delta^2\bigr)} \Delta^2\Bigr)
  -\frac{\bigl(1+\frac{(Q-1) C_1}{Q}\bigr) K}{2 Q}
  \Bigl(\frac{1+\frac{Q^2}{Q-1} \Delta^2}{1+\frac{Q C_1}{Q-1} \Delta^2}
  \Bigr)\ln \Bigl(1-\frac{\frac{Q^3}{(Q-1)^2}}{1+\frac{Q^2}{Q-1} \Delta^2
  } \Delta^2 \Bigr) \; .
\end{eqnarray}

\begin{figure}
  \centering
  \subfigure[]{
    \label{fig:RhoAsym}
    \includegraphics[angle=270, width=0.475\linewidth]{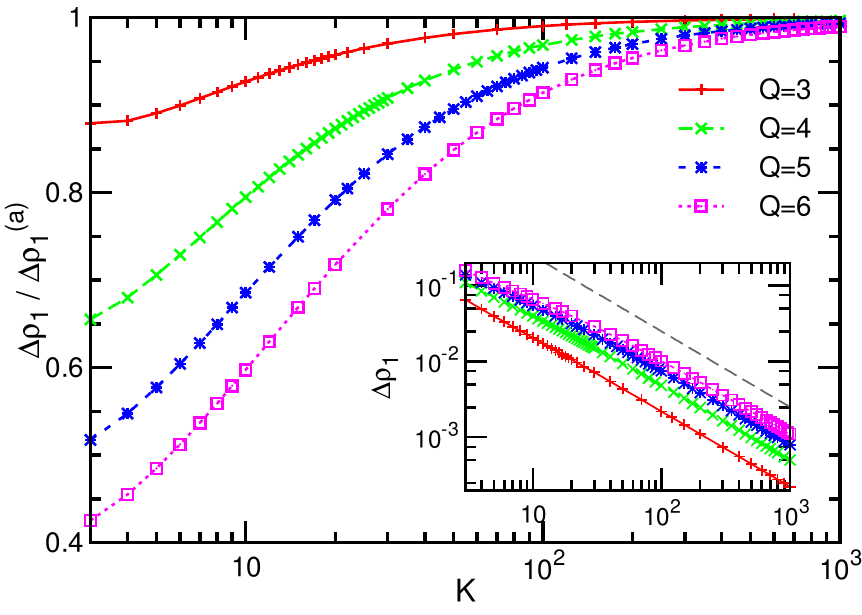}
  }
  \subfigure[]{
    \label{fig:BetaAsym}
    \includegraphics[angle=270, width=0.475\linewidth]{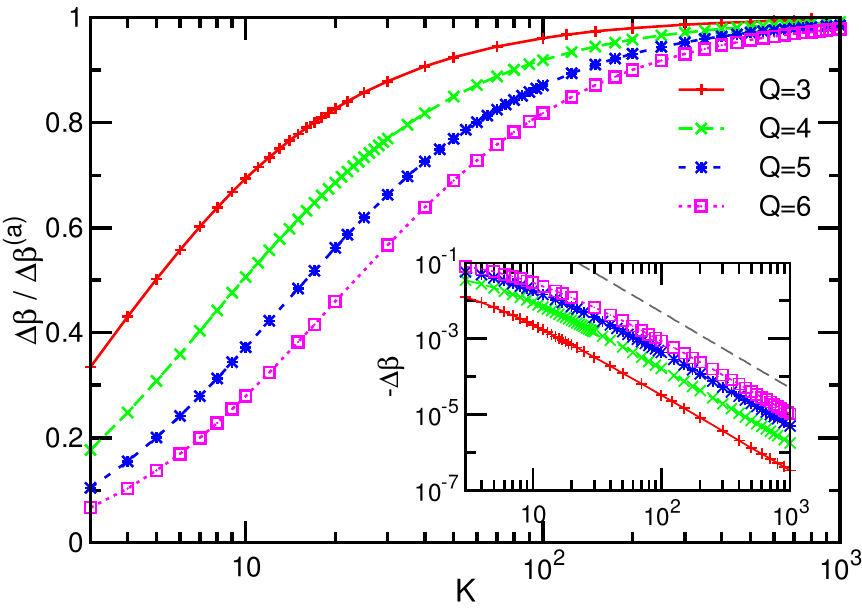}
  }
  \caption{
    \label{fig:Asym}
    Asymptotic behaviors of the gap $\Delta \rho_1$ of the dominant color density (a) and the gap $\Delta \beta$ of the microcanonical inverse temperature (b), at the microcanonical SSB phase transition on RR graphs of degree $K$. The different sets of theoretical curves are for different values of $Q$. In the main panel of (a) $\Delta \rho_1$ is rescaled by $\Delta \rho_1^{(a)} = \frac{(Q-1)(Q-2)}{3 Q} K^{-1}$, while in the main panel of (b) $\Delta \beta$ is rescaled by $\Delta \beta^{(a)} = - \frac{Q (Q-2)^2}{9} K^{-2}$. The insets of (a) and (b) demonstrate the $K^{-1}$ asymptotic decay of $\Delta \rho_1$ and the $K^{-2}$ asymptotic decay of $\Delta \beta$, with the $K^{-1}$ and $K^{-2}$ power laws marked by the two dashed lines.
  }
\end{figure}

If we assume $\Delta$ to be small, then based on the BP equation the expression for $\Delta$ is
\begin{equation}
  \Delta = \frac{2 Q (Q-1)}{(Q-2) C_1^2 (K-1) (K-2)}
  \Bigl(1- \frac{(K-1)C_1}{Q}\Bigr) \; .
\end{equation}
In the limit of $K \rightarrow \infty$, we find that, to fourth order of $\Delta$,
\begin{equation}
  s_{{\rm diff}} = - \Bigl( \frac{Q K C_1}{2 (Q-1)} -
  \frac{K (K-1) C_1^2}{2 (Q-1)}
  \Bigr) \Delta^2 +  \frac{(Q-2) K (K-1) (K-2) C_1^3}{6 (Q-1)^2} \Delta^3 +
  \frac{K Q^4}{4 (Q-1)^3} \bigl(1+\frac{(Q-1)C_1}{Q}\bigr)
  \bigl(1-\frac{C_1}{Q}\bigr) \Delta^4 \; .
\end{equation}
In the limit of $K\rightarrow \infty$, it turns out that $C_1 (K-1)/Q$ is very close to unity, so we write
\begin{equation}
  C_1 = \frac{Q-\varepsilon}{K-1}
\end{equation}
with $\varepsilon$ being a small quantity. To the leading order of $\varepsilon$, we have 
\begin{equation}
  \Delta = \frac{2 (Q-1)}{Q^2 (Q-2)} \varepsilon \; .
\end{equation}
Then we obtain from the condition $s_{{\rm diff}}=0$ that
\begin{equation}
  \Delta = \frac{(Q-1) (Q-2)}{3 K Q} \; , \quad \quad
  \varepsilon = \frac{Q (Q-2)^2}{6 K} \; .
\end{equation}
Because $\rho_1 = \frac{1}{Q} + \Delta$ for $K\rightarrow \infty$, we see that the asymptotic behavior of the jump $\Delta \rho_1$ at $u_{{\rm mic}}$ is
\begin{equation}
  \Delta \rho_1 \approx  \frac{(Q-1) (Q-2)}{3 K Q} \; .
\end{equation}
This asymptotic scaling behavior is in agreement with numerical computations, see Fig.~\ref{fig:RhoAsym}.

At a given energy density $u$, the difference between the microcanonical inverse temperature of the MP and DS phases is
\begin{equation}
  \Delta \beta = -\ln \Bigl(\frac{1+\frac{Q^2}{Q-1} \Delta^2}
         {1-\frac{Q^2}{(Q-1)^2} \Delta^2} \Bigr) \; .
\end{equation}
At the microcanonical SSB phase transition point, the scaling behavior of $\Delta \beta$ is then
\begin{equation}
  \Delta \beta \approx - \frac{Q^3}{(Q-1)^2} \Delta^2 \approx
  - \frac{Q (Q-2)^2}{9 K^2} \; .
\end{equation}
This asymptotic scaling behavior of $\Delta \beta$ is also in agreement with numerical computations, see Fig.~\ref{fig:BetaAsym}.

These asymptotic results suggest that the microcanonical SSB phase transition on RR graphs will be discontinuous for any finite value of degree $K$, and it becomes continuous only at $K=\infty$, i.e., when the graph becomes completely connected.

\section*{S5: The Potts model with large $Q$ values on RR graphs of fixed degree $K$}

\begin{figure}
  \centering
  \subfigure[]{
    \label{fig:RhoAsymQ}
    \includegraphics[angle=270, width=0.475\linewidth]{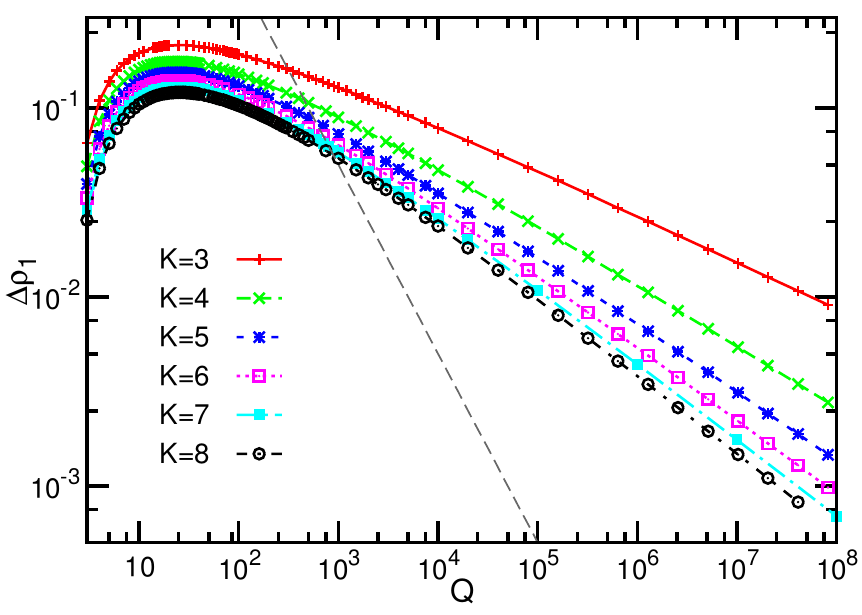}
  }
  \subfigure[]{
    \label{fig:BetaAsymQ}
    \includegraphics[angle=270, width=0.475\linewidth]{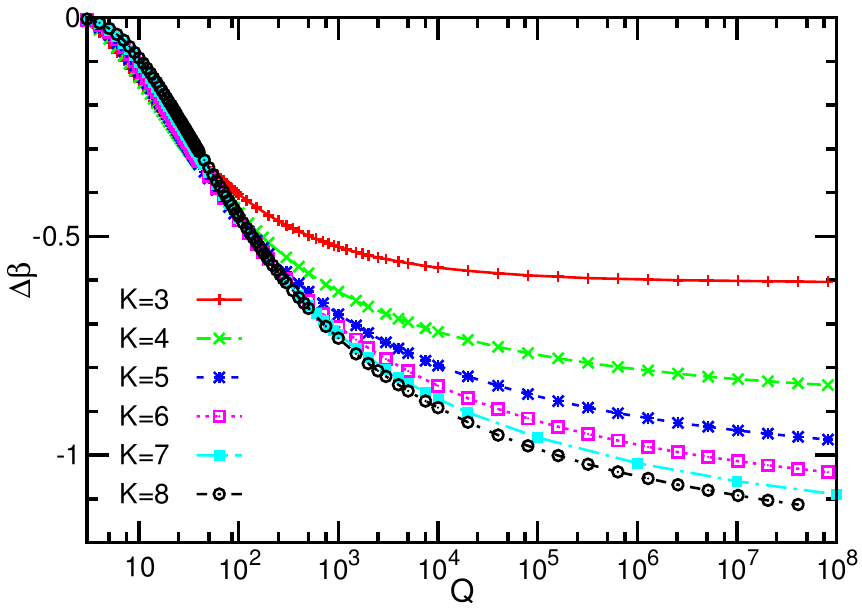}
  }
  \caption{
    \label{fig:AsymQ}
    Asymptotic behaviors of the gap $\Delta \rho_1$ of the dominant color density (a) and the gap $\Delta \beta$ of the microcanonical inverse temperature (b), at the microcanonical SSB phase transition on RR graphs of degree $K$, as the number $Q$ of colors changes. The different sets of theoretical curves are for different values of $K$. The dashed line in (a) marks the scaling behavior $\Delta \rho_1 \propto Q^{-1}$, which is much steeper than the actual decaying behaviors of $\Delta \rho_1$.
  }
\end{figure}

It is also interesting to see how the microcanonical SSB phase transition behaves at the limit of large $Q$. For RR graphs of fixed degree $K$ we can determine the gaps $\Delta \rho_1$ and $\Delta \beta$ at $u_{{\rm mic}}$ as a function of $Q$, see Fig.~\ref{fig:AsymQ}. We observe that $\Delta \rho_1$ is not monotonic in $Q$ but it attains a maximum value at $Q\approx 20$ and then decays slowly as a power law ($\Delta \rho_1 \sim Q^{-\gamma}$) with exponent $\gamma$ much smaller than unity. On the other hand, $\Delta \beta$ is monotonic in $Q$ and approaches a final negative value as $Q\rightarrow \infty$.  A finite value of $\Delta \beta$ at $Q\rightarrow \infty$ is reasonable because the microcanonical inverse temperature is the slope of $s(u)$.

It would also be interesting to know the limiting behavor of the Potts model as both $Q$ and $K$ approach infinity. The results in Fig.~\ref{fig:Asym}(a) and Fig.~\ref{fig:AsymQ}(a) indicate that $\Delta \rho_1$ will decay to zero no matter whether $Q$ approaches infinity faster or slower than $K$. For $\Delta\beta$, as it is expected to have a finite limiting value at $Q\rightarrow \infty$ at each fixed value of $K$ and it decays as $K^{-2}$ at $K\rightarrow \infty$ for each fixed value of $Q$, we conjecture that $\Delta\beta$  will decay to zero as both $Q$ and $K$ approach infinity, and consequently the microcanonical SSB phase transition will become continuous at the limit of $Q$ and $K$ both approach infinity.

\section*{S6: Potts model with kinetic energies}

The Potts model discussed in the main text only considers the interaction energies between neighboring vertices in the graph. Here we introduce local kinetic energies to the vertices to make the model more general~\cite{MartinMayor-2007,Schierz-etal-2016}. Suppose there is a particle of mass $m$ on top of each vertex $i$ and this particle can move in a small confined space so it has a kinetic energy $\frac{\bm{p}_i^2}{2m}$, where $\bm{p}_i$ is the momentum of this particle. The dimensionality of $\bm{p}_i$ is denoted as $D^\prime$. The total energy of this extended system then depends on the color configuration $\bm{c}=(c_1, c_2, \ldots, c_N)$ and the momentum vectors $\{\bm{p}_i\}$ of all the vertices:
\begin{equation}
  E_{{\rm total}} = -\sum\limits_{(i, j)\in G} \delta_{c_i}^{c_j}
  +  \sum\limits_{i=1}^{N} \frac{\bm{p}_i^2}{2 m} \; .
\end{equation}

When the total energy of this system is restricted to a tiny interval $[E, E + \Delta E)$, where $E$ is extensive and $\Delta E \rightarrow 0$, the partition function of the system is
\begin{equation}
  Z = \int\limits_{E}^{E+\Delta E} {\rm d} E_{{\rm total}}
  \sum\limits_{\bm{c}} \prod\limits_{i=1}^{N} \int
  \frac{{\rm d} \bm{p}_i}{p_0^{D^\prime}}
  \delta\Bigl( E_{{\rm total}} - E(\bm{c}) - \sum\limits_{i}
  \frac{\bm{p}_i^2}{2 m}\Bigr)
  \; ,
\end{equation}
where $p_0$ is certain characteristic momentum value needed to count the number of microscopic states in the momentum space, and $E(\bm{c})$ is simply the color energy. By integrating out the momentum degrees of freedom we obtain that
\begin{equation}
  Z  =  \frac{4 m \pi^{N D^\prime /2}}{\Gamma\bigl(\frac{N D^\prime}{2}\bigr)
    p_0^{N D^\prime}} \int\limits_{E}^{E+\Delta E} {\rm d} E_{{\rm total}}
  \sum\limits_{\bm{c}}
  \bigl[2 m (E_{{\rm total}} - E(\bm{c})\bigr]^{\frac{N D^\prime -1}{2}} \; .
\end{equation}

\begin{figure}
  \centering
  \includegraphics[angle=270, width=0.475\linewidth]{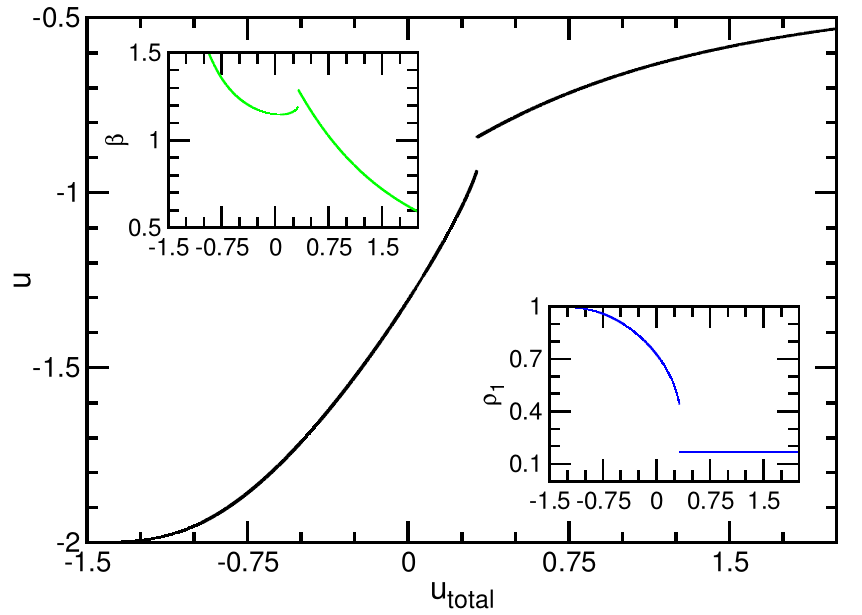}
  \caption{
    \label{fig:Kinetic}
    The Potts model with kinetic energies for regular random graphs ($K\! = \! 4, Q\! = \! 6$, and $D^\prime \! = \! 3$). The discontinuous phase transition occurs at $u_{{\rm total}}\! =\! 0.321$, at which the value of color interaction energy density $u$ drops from $u\! =\! -0.842$ to $u\! =\! -0.936$, the density of the dominant color jump from $\rho_1\! = \! 1/6$ to $\rho_1\! = \! 0.440$, and the microcanonical inverse temperature drops from $\beta\! = \! 1.290$ to $\beta\! = \! 1.193$.
      }
\end{figure}

Let us denote the total energy density of the system as $u_{{\rm total}}$, that is, $E = N u_{{\rm total}}$. By noticing that the total number of color configurations at energy $E(\bm{c}) = N u$ is $\exp\bigl( s(u)\bigr)$, where $s(u)$ is the entropy density of color configurations at given interaction energy density $u$, we can re-write the above expression as
\begin{equation}
  Z \propto \int\limits_{-\infty}^{u_{{\rm total}}} {\rm d} u
  \exp\Bigl( N \bigl( s(u) + \frac{D^\prime}{2} \ln \frac{u_{{\rm total}}-u}{\varepsilon_0}\bigr)\Bigr) \; ,
  \label{eq:Ztotal}
\end{equation}
where $\varepsilon_0 \equiv \frac{p_0^2}{2 m}$. From Eq.~(\ref{eq:Ztotal}) we see that at a given value of total energy density $u_{{\rm total}}$, the mean interaction energy density $u^*$ of the system will be determined by
\begin{equation}
  u^* =\underset{u}{\arg\max}
  \Bigl( s(u) + \frac{D^\prime}{2} \ln \frac{u_{{\rm total}} - u}{\varepsilon_0}
  \Bigr) \; .
\end{equation}
Therefore $u^*$ must be a root of
\begin{equation}
  u_{{\rm total}} - u^* = \frac{D^\prime}{2} \frac{1}{\beta(u^*)} \; ,
  \label{eq:ustar}
\end{equation}
where $\beta(u) \equiv \frac{{\rm d} s(u)}{{\rm d} u}$ is the microcanonical inverse temperature of the system. Notice that this equilibrium interaction energy density $u^*$ does not depend on $\varepsilon_0$.

Equation (\ref{eq:ustar}) simply says that the mean kinetic energy of a vertex is equal to $\frac{D^\prime}{2}\frac{1}{\beta(u^*)}$. For intermediate values of $u_{{\rm total}}$, Eq.~(\ref{eq:ustar}) has a pair of solutions $u^*$, and the one which corresponds to higher total entropy density will be the physically relevant solution. As demonstrated in Fig.~\ref{fig:Kinetic} for regular graphs ($K\! =\! 4, Q\! = \! 6$, and momentum dimensionality setting to be $D^\prime\! =\! 3$), there is a discontinuous phase transition when the total energy density is decreased to the critical value $u_{{\rm total}}\! = \! 0.321$. Such a discontinuous phase transition will occur for other values of $K\! \geq \! 3$ and $Q\! \geq \! 3$ as well.

\section*{S7: Bond-diluted lattice systems and short-range interaction range $l$}

The vertices in a $D$-dimensional hypercubic lattice of side length $L$ are located at positions $(x_1, x_2, \ldots, x_D)$ where $x_d$ ($d=1, 2, \ldots, D$) are integer values.  Periodic boundary conditions are imposed, so that $(x_1, x_2, \ldots, x_D)$ and $x_1^\prime, x_2^\prime, \ldots, x_D^\prime)$ are the same position if
\begin{equation}
  (x_d\mod L) = (x_d^\prime\mod L)
\end{equation}
for every $d=1, 2, \ldots, D$. The total number of vertices in the system is $N=L^D$. Each vertex has $2 D$ bonds linking itself to its nearest neighboring vertices in space. The length of the shortest loops in such a hypercubic graph is equal to four and it does not increase with system size $L$. To make it more difficult for nucleation to occur (see the next section), we dilute this hypercubic graph by deleting a large fraction of bonds and keeping only $K$ bonds for each vertex. A maximally random bond-diluted lattice graph is constructed according to the following procedure:
\begin{enumerate}
\item[1.] Construct an initial $D$-dimensional bond-diluted hypercubic lattice graph in which every vertex has exactly $K$ ``active'' bonds (the remaining $2 D - K$ bonds of this vertex are all regarded as ``inactive'').
\item[2.] Pick a vertex $i$ uniformly at random from the lattice graph and pick uniformly at random an active bond $(i, j)$ from its $K$ active bonds; then move to vertex $j$ and pick uniformly at random an inactive bond $(j, k)$ from its $2 D- K$ inactive bonds; then move to vertex $k$ and pick uniformly at random an active bond $(k, l)$ from its $K$ active bonds; ...... This chain of alternative active and inactive bonds is further extended until it visits a vertex that is already in the chain (loop closure).
\item[3.] If the length of this sampled loop is odd, nothing is changed. But if the length of this loop is even, then all the active bonds in this loop are deleted from the graph (i.e., they change to be inactive) while all the originally inactive bonds of this loop are added to the graph (i.e., they now become active). This switching action keeps the ``active'' degree of every involved vertex unchanged.
\item[4.] Repeat steps (2) and (3) a large number of times (e.g., $10000\times N$) to make the bond-diluted lattice graph as random as possible.
\end{enumerate}
This loop-switching algorithm is similar to the algorithm used in Ref.~\cite{Fernandez-etal-2010}. It is easy to prove that this algorithm is ergodic and it leads to a uniform distribution among all the valid $K$-regular lattice graphs. Because all the bonds in the original hypercubic lattice are between spatial nearest neighbors, the constructed bond-diluted graphs naturally contain only bonds between nearest neighbors.

We also consider diluted lattice graphs with longer interaction ranges to explore the effect of interaction range to the microcanonical SSB transition. For this purpose, we consider a lattice system in which each vertex $i$ at position $(x_1, x_2, \ldots, x_D)$ of the periodic hypercubic lattice has $(2 l +1)^D - 1$ bonds to all the other vertices located in or at the surface of the hypercubic box of side length $(2 l+1)$ centered on vertex $i$. That is, there is a bond between position $(x_1, x_2, \ldots, x_D)$ and the positions $(x_1 + \delta x_1, x_2 + \delta x_2, \ldots, x_D + \delta x_D)$ where
\begin{equation}
  \delta x_d \in \{ -l, -l+1, \ldots, -1, 0, 1, \ldots, l-1,  l \}
  \quad \quad \quad (d = 1, 2, \ldots, D) \; .
\end{equation}
This lattice graph is much more densely connected than the simplest hypercubic graph of degree $2 D$, since each vertex has $(2 l+1)^D -1$ attached edges. To make it sparse we only retain $K$ edges for each vertex and delete all the other edges. Such a maximally random diluted graph can be sampled by the same loop-switching algorithm as described above.

\begin{figure}
  \centering
  \includegraphics[angle=270, width=0.475\linewidth]{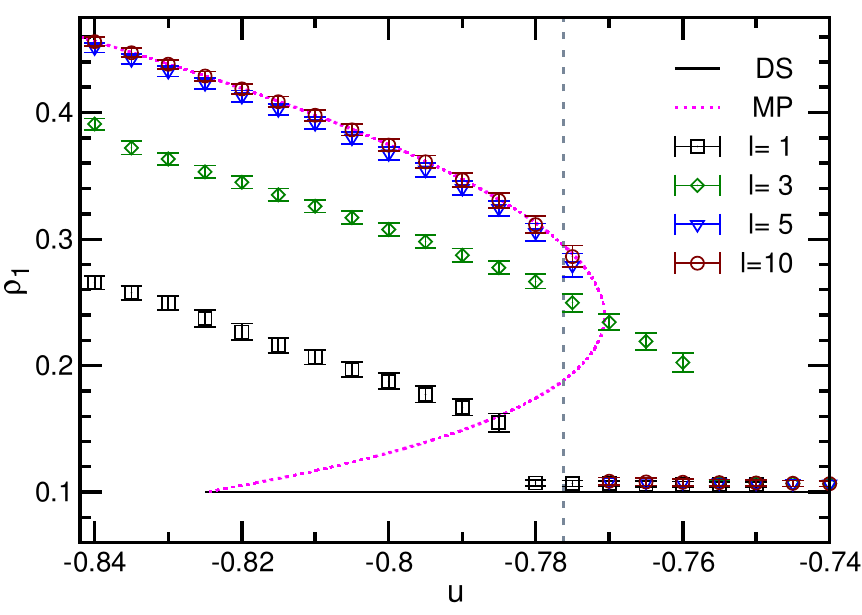}
  \caption{
    \label{fig:3Dl}
    The Potts model ($Q=10$) on four graph instances of three-dimensional short-range interaction lattice systems. Symbols are microcanonical MC simulation results. The side length of the diluted lattice graphs is $L=40$ so the total number of vertices is $N=64000$. Each vertex has $K=3$ attached edges in these graphs, with the interaction range being $l=1$ (squares), $l=3$ (diamonds), $l=5$ (triangle), and $l=10$ (circles). The theoretical predictions for RR graphs of degree $K=3$ concerning the density $\rho_1$ of the dominant color are the solid line (the DS phase) and the dotted line (the MP phase). The vertical dashed line marks the predicted microcanonical SSB phase transition point $u_{{\rm mic}}=-0.776$ for RR graphs of degree $K=3$.
  }
\end{figure}

We have performed some preliminary computer simulations for the physically relevant dimension $D=3$. The side length of the periodic cubic lattice is fixed to $L=40$, the vertex degree is fixed to $K=3$, while four different values are tried for the interaction range parameter $l$, namely $l=1, 3, 5, 10$. The microcanonical MC simulation results shown in Fig.~\ref{fig:3Dl} clearly demonstrate that the interaction range has a dramatic effect on the SSB transition. When the interaction range $l\geq 5$ the simulation results on these finite-size lattice graphs are very similar to the predicted results for random RR graphs.

\section*{S8: Droplet nucleation and phase separation}

Here we review some of the key ideas of the droplet nucleation theory and discuss when droplet formation will be severely suppressed. To be concrete, we consider the $D$-dimensional hypercubic lattice of side length $L$ with periodic boundary conditions. The total number of vertices in the lattice system is $N=L^D$, and the total number of bonds (edges) is $M=D N$ if every vertex only interacts with its $2 D$ nearest neighbors.

In the canonical statistical ensemble and at the thermodynamic limit $L\!\rightarrow\! \infty$, the $Q$-state Potts model defined on such a lattice will experience an equilibrium phase transition at certain critical inverse temperature $\beta_{{\rm c}}$, between the disordered symmetric (DS) phase and the canonical polarized (CP) phase. In the DS phase all the $Q$ colors are equally abundant, while in the CP phase one randomly picked color is favored over all the other colors. We consider the case of this canonical SSB phase transition being discontinuous. Let us denote the energy densities of these two phases at $\beta_{{\rm c}}$ as $u_{{\rm DS}}^{{\rm c}}$ and $u_{{\rm CP}}^{{\rm c}}$, respectively. Similarly, the entropy densities of these two phases at $\beta_{{\rm c}}$ are denoted as $s_{{\rm DS}}^{{\rm c}}$ and $s_{{\rm CP}}^{{\rm c}}$. Because the DS--CP phase transition is an equilibrium one, we have
\begin{equation}
  \frac{s_{{\rm DS}}^{{\rm c}} - s_{{\rm CP}}^{{\rm c}}}{u_{{\rm DS}}^{{\rm c}} - u_{{\rm CP}}^{{\rm c}}} = \beta_{{\rm c}} \; .
\end{equation}

Now we consider the microcanonical ensemble of fixed energy density $u$ and assume $u$ takes an intermediate value between $u_{{\rm CP}}^{{\rm c}}$ and $u_{{\rm DS}}^{{\rm c}}$. Because of the existence of the equilibrium canonical DS--CP phase transition, the entropy density $s(u)$ at this intermediate energy density must satisfy the following inequality:
\begin{equation}
  \label{eq:Senvelop}
  s(u) \leq   r s_{{\rm CP}}^{{\rm c}} + (1-r) s_{{\rm DS}}^{{\rm c}} \; ,
\end{equation}
where the parameter $r$ is defined by
\begin{equation}
  \label{eq:Rfrac}
  r \equiv \frac{u_{{\rm DS}}^{{\rm c}} - u}{u_{{\rm DS}}^{{\rm c}}
    - u_{{\rm CP}}^{{\rm c}}}  \; .
\end{equation}
At the thermodynamic limit $L\! \rightarrow\! \infty$ it turns out that $s(u)$ achieves the upper-limit of the inequality (\ref{eq:Senvelop}) by phase separation. The argument goes as follows. Suppose the DS and CP phases coexist in the system and $r$ is the relative size of the CP phase. In the case of $r$ close to zero, to minimize the surface area between these two phases, we may assume that the CP phase is confined within a hyperspherical droplet of radius $R$. The total number $n_{V}$ of vertices in this droplet is then $n_V\! = \! c_0 R^D$ while the total number $n_{S}$ of vertices on its surface is $n_S\! = \! c_1 R^{D-1}$, where $c_0, c_1$ are two constants. When the droplet becomes large in size, we see that
\begin{equation}
  \lim\limits_{R \rightarrow \infty} \frac{n_S}{n_V} = 0 \; .
  \label{eq:SvsVcond}
\end{equation}
A consequence of Eq.~(\ref{eq:SvsVcond}) is that, when $R$ becomes sufficiently large the surface interaction energy between the CP and DS phases will be negligible in comparison with the volume interaction energy of the droplet. Then the CP droplet and the DS subsystem can be treated as two independent systems and the entropy density of the combined system is then
\begin{equation}
  s(u) = r s_{{\rm CP}}^{{\rm c}} + (1-r) s_{{\rm DS}}^{{\rm c}} \; .
  \label{eq:Slimit}
\end{equation}
Because of Eq.~(\ref{eq:Rfrac}), the first derivative of $s(u)$ at $u\! =\! u_{{\rm DS}}^{{\rm c}}$ is equal to $\beta_{{\rm c}}$. Consequently $s(u)$ is $C^1$-continuous at $u_{{\rm DS}}^{{\rm c}}$. According to this droplet picture, in the thermodynamic limit $L\! \rightarrow\! \infty$, when the energy density $u$ is decreased to $u_{{\rm DS}}^{{\rm c}}$, phase separation starts to occur and the relative size $r$ of the CP droplet increases gradually from $r\! =\! 0$ at $u\! =\! u_{{\rm DS}}^{{\rm c}}$ to $r\! =\! 1$ as $u$ is gradually decreased to $u\! =\! u_{{\rm CP}}^{{\rm c}}$. The inverse temperature of the system keeps the value $\beta_{{\rm c}}$ during this CP phase expansion process. Since $r$ continuously increases from zero, the density $\rho_1$ of the dominant color must also deviate from $1/Q$ in a continuously manner.

For a stable droplet to form in the system, however, the size $R$ of the droplet must exceed certain threshold length $R_{{\rm th}}(u)$, which depends on the energy density $u$ of the whole system. If $R$ is too small, the energy (and free energy) gain of forming a partially ordered droplet will not be enough to compensate for the penalty of interfacial energy, and then the droplet will be suppressed. If the side length $L$ of the system is comparable or even smaller than $R_{{\rm th}}(u)$, then it is likely the system as whole will change directly from the DS phase to a partially ordered phase, without experiencing the intermediate phase coexistence stage.  This partially ordered phase for such finite-size systems may contain a percolating cluster of connected vertices which are all in the dominant color state (see, for example, Ref.~\cite{MacDowell-etal-2006} for related simulation results obtained on the supercooled liquid/gas system).

The required minimum radius $R_{{\rm th}}$ for droplet formation and phase separation might be greatly increased in a bond-diluted lattice system as compared to an intact lattice system. When bonds are deleted in a maximally random manner and the active degree $K$ of every vertex is quite small, the graph will be locally quite similar to a random graph, and the typical length $\ell_{{\rm loop}}$ of the shortest path of unbroken bonds linking two spatial neighbors will be relatively large. Indeed, as the spatial dimensionality $D$ of the lattice increases while the active degree $K$ of the vertices is fixed, short loops will be more and more difficult to form and the graph will be more and more like a completely random graph. If the radius $R$ of a hyperspherical region in such a bond-diluted spatial graph is of the same order as $\ell_{{\rm loop}}$, the vertices of this region will be well approximated by a tree and its surface interaction energy will be comparable to the volume interaction energy of this region. Therefore, for phase separation to occur the side length $L$ of the lattice system must be much larger than $\ell_{{\rm loop}}$. 

Now we offer a rough estimate of $\ell_{{\rm loop}}$. Consider a rooted tree in the bond-diluted hypercubic lattice and assume the path length between two leaf vertices is $L_{{\rm tree}}$ (the distance from the root to a leaf vertex is $L_{{\rm tree}}/2$). Since each internal vertex of this tree has $K$ attached edges, the total number of vertices in the tree is approximately $K^{L_{{\rm tree}}/2}$. Because the directions of the edges in this tree are quite random, the length of the spatial region which contains this tree is approximately $\sqrt{L_{{\rm tree}}}$, and the total number of vertices of this region is then approximately $\bigl(\sqrt{L_{{\rm tree}}}\bigr)^D$. To guarantee the absence of loops the total number of vertices in the tree must not exceed the allowed number of vertices in the region. Therefore, we estimate $\ell_{{\rm loop}}$ to be the largest integer value $L_{{\rm tree}}$ which satisfies the following condition:
\begin{equation}
  K^{L_{{\rm tree}}/2}  \leq \bigl(L_{{\rm tree}}\bigr)^{D/2} \; .
\end{equation}

In the case of $K=4$ and $D=8$ as in Fig.~2(d), the above condition suggests that $\ell_{{\rm loop}} = 16$, which indicates a threshold number of vertices exceeding $N_{{\rm th}}=(\ell_{{\rm loop}})^D = 4.3\times 10^{9}$ (which is much larger than the accessible graph sizes in our computer simulations).  If the interaction range $l$ in the lattice system increases while the vertex degree $K$ keeps fixed (see the preceding section), because the Euclidean length of an edge of this system is approximately $l$, the above condition probably needs to be modified as
\begin{equation}
  K^{L_{{\rm tree}}/2}  \leq l^D \bigl(L_{{\rm tree}}\bigr)^{D/2} \; ,
\end{equation}
and the characteristic length $\ell_{{\rm loop}}$ will increase with $l$. For the $D=3$ graph instances of Fig.~\ref{fig:3Dl}, the estimated lengths are $\ell_{{\rm loop}}=14$ for $l=3$ (the lower-bound of threshold number of vertices is then $N_{{\rm th}} = (l \ell_{{\rm loop}})^D= 7.4\times 10^4$), $\ell_{{\rm loop}}=17$ for $l=5$ ($N_{{\rm th}}=6.1\times 10^5$), and $\ell_{{\rm loop}}=21$ for $l=10$ ($N_{{\rm th}}=9.3\times 10^6$). These quantitative estimates may help explain why at $l\geq 5$ the simulation results of Fig.~\ref{fig:3Dl} for the $D=3$ spatial graphs are quite close to the theoretical results predicted for RR graphs.

\end{appendix}


\begin{thebibliography}{10}

\bibitem{Brading-etal-2017}
  K.~Brading, E.~Castellani, and N.~Teh,
  \newblock Symmetry and symmetry breaking,
  \newblock In {\em The Stanford Encyclopedia of
    Philosophy} (Metaphysics Research Lab, Stanford University, winter 2017
  edition, 2017), edited by E.~N. Zalta.
  
\bibitem{Potts-1952}
  R.~B. Potts,
  \newblock Some generalized order-disorder transformations,
  \newblock Proc. Cambridge Phil. Soc. {\bf 48}, 106 (1952).
  
\bibitem{Wu-1982}
  F.~Y. Wu,
  \newblock The Potts model,
  \newblock Rev. Mod. Phys. {\bf 54}, 235 (1982).
  
\bibitem{Baxter-1982}
  R.~J. Baxter,
  \newblock {\em Exactly Solved Models in Statistical Mechanics}
  (Academic Press, London, UK, 1982).
  
\bibitem{Gorbenko-etal-2018}
  V.~Gorbenko, S.~Rychkov, and B.~Zan,
  \newblock Walking, weak first-order transitions, and complex CFTs II.
  Two-dimensional Potts model at $q>4$,
  \newblock SciPost Phys. {\bf 5}, 050 (2018).
  
\bibitem{Blote-etal-2017}
  H.~W.~J. Bl\"ote, W.~Guo, and M.~P. Nightingale,
  \newblock Scaling in the vicinity of the four-state Potts fixed point,
  \newblock J. Phys. A: Math. Theor. {\bf 50}, 324001 (2017).
  
\bibitem{Hu-Deng-2015}
  H.~Hu and Y.~Deng,
  \newblock Universal critical wrapping probabilities in the canonical ensemble,
  \newblock Nuclear Phys. B {\bf 898}, 157 (2015).

\bibitem{Wang-Xie-etal-2014}
  S.~Wang, Z.-Y. Xie, J.~Chen, B.~Normand, and T.~Xiang,
  \newblock Phase transitions of ferromagnetic Potts model on the simple cubic
  lattice,
  \newblock Chinese Phys. Lett. {\bf 31}, 070503 (2014).
  
\bibitem{Lee-Lucas-2014}
  C.~H. Lee and A. Lucas,
  \newblock Simple model for multiple-choice collective decision making,
  \newblock Phys. Rev. E {\bf 90} 052804 (2014).
  
\bibitem{Tian-etal-2013}
  L.~Tian, H.~Ma, W.~Guo, and L.-H. Tang,
  \newblock Phase transitions of the $q$-state Potts model on multiply-laced
  Sierpinski gaskets,
  \newblock Eur. Phys. J. B {\bf86}, 197 (2013).
  
\bibitem{Chen-etal-2011}
  Q.~N. Chen, M.~P. Qin, J.~Chen, Z.~C. Wei, H.~H. Zhao, B.~Normand, and
  T.~Xiang,
  \newblock Partial order and finite-temperature phase transitions in Potts
  models on irregular lattices,
  \newblock Phys. Rev. Lett. {\bf 107}, 165701 (2011).
  
\bibitem{Deng-etal-2011}
  Y.~Deng, Y.~Huang, J.~L. Jacobsen, J.~Salas, and A.~D. Sokal,
  \newblock Finite-temperature phase transition in a class of four-state Potts
  antiferromagnets,
  \newblock Phys. Rev. Lett. {\bf 107}, 150601 (2011).
  
\bibitem{Gross-etal-1996}
  D.~H.~E. Gross, A.~Ecker, and X.~Z. Zhang,
  \newblock Microcanonical thermodynamics of first order phase transitions
  studied in the Potts model,
  \newblock Ann. Physik {\bf 508}, 446 (1996).
  
\bibitem{Gross-1997}
  D.~H.~E. Gross,
  \newblock Microcanonical thermodynamics and statistical fragmentation of
  dissipative systems: The topological structure of the $n$-body phase space,
  \newblock Phys. Rep. {\bf 279}, 119 (1997).

\bibitem{MartinMayor-2007}
  V.~{Martin-Mayor},
  \newblock Microcanonical approach to the simulation of first-order phase
  transitions,
  \newblock Phys. Rev. Lett. {\bf 98}, 137207 (2007).
  
\bibitem{Moreno-etal-2018}
  F.~Moreno, S.~Davis, C.~Loyola, and J.~Peralta,
  \newblock Ordered metastable states in the Potts model and their connection
  with the superheated solid state,
  \newblock Physica A {\bf 509}, 361 (2018).
  
\bibitem{SInote2019}
  See Supplementary Information for additional theoretical and numerical results, some technical details on constructing a bond-diluted lattice system, and a qualitative discussion of the nucleation phenomenon in finite-dimensional systems.

\bibitem{Biskup-etal-2002}
  M.~Biskup, L.~Chayes, and R.~Koteck\'y,
  \newblock On the formation/dissolution of equilibrium droplets,
  \newblock Europhys. Lett. {\bf 60}, 21 (2002).
  
\bibitem{Binder-2003}
  K.~Binder,
  \newblock Theory of the evaporation/condensation transition of equilibrium
  droplets in finite volumes,
  \newblock Physica A {\bf 319}, 99 (2003).
  
\bibitem{MacDowell-etal-2006}
   L.~G.~MacDowell, V.~K. Shen, and J.~R. Errington,
    \newblock Nucleation and cavitation of spherical, cylindrical, and
    slablike droplets and bubbles in small systems,
    \newblock J. Chem. Phys. {\bf 125}, 034705 (2006).
  
\bibitem{Nogawa-etal-2011}
  T.~Nogawa, N.~Ito, and H.~Watanabe,
  \newblock Evaporation-condensation transition of the two-dimensional Potts
  model in the microcanonical ensemble,
  \newblock Phys. Rev. E {\bf 84}, 061107 (2011).
  
\bibitem{Xu-etal-2018}
  Y.-Z. Xu, C.~H. Yeung, H.-J. Zhou, and D.~Saad,
  \newblock Entropy inflection and invisible low-energy states: Defensive
  alliance example,
  \newblock Phys. Rev. Lett. {\bf 121}, 210602 (2018).
  
\bibitem{Mezard-Montanari-2009}
  M.~M{\'{e}}zard and A.~Montanari, 
  \newblock {\em Information, Physics, and Computation} (Oxford Univ. Press,
  New York, 2009).
  
\bibitem{Mukamel-2008}
  D.~Mukamel,
  \newblock Statistical mechanics of systems with long range interactions,
  \newblock AIP Conf. Proc. {\bf 970}, 22 (2008).
  
\bibitem{Campa-etal-2009}
  A.~Campa, T.~Dauxois, and S.~Ruffo,
  \newblock Statistical mechanics and dynamics of solvable models with
  long-range interactions,
  \newblock Phys. Rep. {\bf 480}, 57 (2009).

  \bibitem{Murata-Nishimori-2012}
    Y.~Murata and H.~Nishimori,
    \newblock Ensemble inequivalence in the spherical spin glass model with
    nonlinear interactions,
    \newblock J. Phys. Soc. Jpn. {\bf 81}, 114008 (2012).
    
  \bibitem{Touchette-2015}
    H.~Touchette,
    \newblock Equivalence and nonequivalence of ensembles: Thermodynamic,
    macrostate, and measure levels,
    \newblock J. Stat. Phys. {\bf 159}, 987 (2015).

  \bibitem{Huang-1987}
    K.~Huang,
    \newblock {\em Statistical Mechanics} (John Wiley, New York,
    second edition, 1987).
    
    
  \bibitem{Mezard-etal-1987}
    M.~M{\'{e}}zard, G.~Parisi, and M.~A. Virasoro,
    \newblock {\em Spin Glass Theory and Beyond}
    (World Scientific, Singapore, 1987).
    
  \bibitem{Xiao-Zhou-2011}
    J.-Q. Xiao and H.-J. Zhou,
    \newblock Partition function loop series for a general graphical model:
    free-energy corrections and message-passing equations,
    \newblock J. Phys. A: Math. Theor. {\bf 44}, 425001 (2011).
    
  \bibitem{Zhou-Wang-2012}
    H.-J. Zhou and C.~Wang,
    \newblock Region graph partition function expansion and approximate free
    energy landscapes: Theory and some numerical results,
    \newblock J. Stat. Phys. {\bf 148}, 513 (2012).

  \bibitem{Pearl-1988}
    J.~Pearl,
    \newblock {\em Probabilistic Reasoning in Intelligent Systems: Networks of
      Plausible Inference} (Morgan Kaufmann, San Franciso, CA, USA, 1988).


\bibitem{Creutz-1983}
  M.~Creutz,
  \newblock Microcanonical Monte Carlo simulation,
  \newblock Phys. Rev. Lett. {\bf 50}, 1411 (1983).
  
\bibitem{Lee-1995}
  K.-C. Lee,
  \newblock Rejection-free Monte Carlo technique,
  \newblock J. Phys. A: Math. Gen. {\bf 28}, 4835 (1995).
  
\bibitem{Schierz-etal-2016}
  P.~Schierz, J.~Zierenberg, and W.~Janke,
  \newblock First-order phase transitions in the real microcanonical ensemble,
  \newblock Phys. Rev. E {\bf 94}, 021301(R) (2016).
  
\bibitem{Fernandez-etal-2010}
   L.~A. Fern\'andez,  V.~Martin-Mayor, G.~Parisi, and B. Seoane,
    \newblock Spin glasses on the hypercube,
    \newblock Phys. Rev. B {\bf 81}, 134403 (2010).
  
\bibitem{Zhou-Ma-2009}
  H.-J. Zhou and H.~Ma,
  \newblock Communities of solutions in single solution clusters of a random
  $k$-satisfiability formula,
  \newblock Phys. Rev. E {\bf 80}, 066108 (2009).
  
\bibitem{Zhou-Wang-2010}
  H.-J. Zhou and C.~Wang,
  \newblock Ground-state configuration space heterogeneity of random
  finite-connectivity spin glasses and random constraint satisfaction problems,
  \newblock J. Stat. Mech.: Theor. Exp. {\bf 2010}, P10010 (2010).
  
\end{thebibliography}
\end{document}